\documentclass[aps, pre, reprint, superscriptaddress, floatfix]{revtex4-2}
\usepackage[english]{babel}
\usepackage[utf8]{inputenc}
\usepackage[T1]{fontenc}
\usepackage{hyperref, xcolor, graphicx}
\usepackage{amsfonts, amssymb, amsmath, mathtools}

\newcommand\beq{\begin{equation}}
\newcommand\eeq{\end{equation}}
\newcommand\bem{\begin{pmatrix}}
\newcommand\eem{\end{pmatrix}}
\newcommand{\tr}{\mathrm{tr}}
\newcommand{\pd}{\partial}
\newcommand{\mO}{\mathcal{O}}
\newcommand{\mH}{\mathcal{H}}

\newcommand{\mU}{\mathcal{U}}
\newcommand{\la}{\langle}
\newcommand{\ra}{\rangle}

\hypersetup{
    colorlinks=true,
    linkcolor=black,
    citecolor=blue,   
    urlcolor=blue,
}

\begin{document}

\title{Streamlined Krylov construction and classification of ergodic Floquet systems}

\author{Nikita Kolganov}
    \affiliation{Institute for Theoretical and Mathematical Physics, Moscow State University, Leninskie Gory, GSP-1, Moscow, 119991, Russia}
\author{Dmitrii~A.~Trunin}
    \email{dmitrii.trunin@princeton.edu}
    \affiliation{Department of Physics, Princeton University, Princeton, NJ 08544, USA}

\date{\today}

\begin{abstract}
We generalize Krylov construction to periodically driven (Floquet) quantum systems using the theory of orthogonal polynomials on the unit circle. Compared to other approaches, our method works faster and maps any quantum dynamics to a one-dimensional tight-binding Krylov chain, which is efficiently simulated on both classical and quantum computers. We also suggest a classification of chaotic and integrable Floquet systems based on the asymptotic behavior of Krylov chain hopping parameters (Verblunsky coefficients). We illustrate this classification with random matrix ensembles, kicked top, and kicked Ising chain.
\end{abstract}

\maketitle

\textit{\textbf{Introduction---}} The emergence of ergodic behavior in isolated quantum systems is an old but thriving subject. Although most local interacting systems were believed to be fully ergodic, i.e., obey strong Eigenstate Thermalization Hypothesis~\cite{Deutsch:1991,Srednicki:1994,DAlessio:2015}, recent studies revealed several notable exceptions from this rule. In particular, such novel ergodicity breaking scenarios encompass many-body localization~\cite{Nandkishore:2014,Altman:2014,Abanin:2018}, quantum many-body scars~\cite{Serbyn:2020,Chandran:2022,Moudgalya:2021}, and Hilbert space fragmentation~\cite{Moudgalya:2021,Sala:2020,Khemani:2019,Motrunich:2021}. Besides, the emergence of ergodic behavior is closely related to quantum chaos, which has recently become increasingly important for understanding quantum field theory and quantum gravity~\cite{MSS,Swingle:2018,Xu:2024,Parker:2018,Balasubramanian:2022-1,Dymarsky:2024}. Therefore, it is crucial to develop methods that capture the dynamics of ergodic and near-ergodic quantum systems in the most efficient way.

Krylov construction is one of the most powerful tools to study quantum dynamics~\cite{Parker:2018,Balasubramanian:2022-1,Dymarsky:2024,Viswanath:book}. This construction maps an arbitrary continuous time evolution into a one-dimensional local hopping problem, which allows one to identify the minimal subspace where time evolution unfolds and follow this dynamics cost-effectively. The efficiency of Krylov construction explains its numerous applications, including quantum computing and control~\cite{Motta:2019,Stair:2019,Cortes:2022,Larocca:2021,Takahashi:2023}, classification of topological phases~\cite{Yates:2020,Caputa:2022-1,Caputa:2022-2,Bhattacharjee:2022-MBL}, semiclassical chaos~\cite{Bhattacharjee:2022,Hashimoto:2023}, and holography~\cite{Dymarsky:2021,Dymarsky:2022,Rabinovici:2019,Rabinovici:2020,Rabinovici:2022,Rabinovici:2023,Jian:2020,Iizuka:2023,Avdoshkin:2019,Caputa:2021,Gorsky:2024,Das:2024}.

In this Letter, we extend Krylov construction to discrete time evolution, i.e., to periodically driven (Floquet) systems. Our algorithm exploits the theory of orthogonal polynomials on the unit circle~\cite{Szego,Simon,Ismail:2005,Ismail:2020,Muck:2022}, works as fast as Lanczos algorithm, and maps \textit{any} discrete time evolution to a one-dimensional local model. Other existing approaches either lack these advantages~\cite{Yates:2021b,Yates:2021a,Yates:2021a,Qi:2024,Nizami:2023,Nizami:2024,Suchsland:2023,Scialchi:2024} or substantially restrict the set of initial states~\cite{Yeh:2023,Yeh:2024} (but not operators~\footnote{The approach of~\cite{Yeh:2023,Yeh:2024} relies on the peculiarities of operator evolution and always generates real Verblunsky coefficients; so, it constructs a Krylov basis for any operator evolution, but not for any state evolution.}). Furthermore, we show that hopping parameters of integrable and chaotic Krylov chains have different asymptotic behavior resembling the universal growth hypothesis~\cite{Parker:2018}. In the limit of infinitely large Hilbert space, this difference implies Krylov localization for integrable chains and delocalization for chaotic ones.

\textit{\textbf{Lanczos algorithm---}} First of all, we briefly review the standard Krylov construction~\cite{Parker:2018,Dymarsky:2024,Balasubramanian:2022-1,Viswanath:book}. Consider a quantum system with a constant Hamiltonian $H$ and $d$-dimensional Hilbert space endowed with an inner product $(A | B) = \left\la A | B \right\ra$ and norm $\| A \| = \sqrt{(A | A)}$. The dynamics of a seeding state $| \psi \ra$ unfolds in the corresponding $D$-dimensional Krylov subspace ($D \le d$):
\beq \label{eq:subspace-L}
|\psi(t)\ra \in \mathrm{span}\!\left\{ | \psi \ra, H | \psi \ra, H^2 | \psi \ra, \cdots, H^{D-1} | \psi \ra \right\}. \eeq
A complete orthonormal basis $\left\{ | \Psi_n \ra \right\}$ in this Krylov subspace is recursively generated by the Lanczos algorithm~\cite{Parker:2018,Dymarsky:2024,Balasubramanian:2022-1,Viswanath:book}:
\beq \label{eq:Lanczos}
\begin{gathered}
    a_n = \la \Psi_n | H | \Psi_n\ra, \qquad
    b_n = \la \Psi_n | H | \Psi_{n-1} \ra, \\
    b_{n+1} | \Psi_{n+1} \ra = H | \Psi_n \ra - a_n | \Psi_n \ra - b_n | \Psi_{n-1} \ra,
\end{gathered} \eeq
with initial conditions $| \Psi_0 \ra = | \psi \ra / \| \psi \|$ and $b_0 = 0$ until $b_D = 0$. This algorithm requires $\mO\!\left(D d^2\right)$ operations and generally works faster than direct diagonalization of the Hamiltonian. Moreover, the Hamiltonian acquires a tri-diagonal form in the basis $\{| \Psi_n \ra \}$:
\beq \label{eq:three-diagonal}
\left\la \Psi_m | H | \Psi_n \right\ra = a_m \delta_{m,n}+ b_m \delta_{m,n+1} + b_n \delta_{m,n-1}. \eeq
In other words, Lanczos algorithm maps the dynamics of a seeding state $| \psi \ra$ into a local one-dimensional hopping problem on the Krylov chain:
\beq \label{eq:tight-binding-Lanczos}
\pd_t \phi_n(t) = i a_n \phi_n(t) - b_{n+1} \phi_{n+1}(t) + b_n \phi_{n-1}(t), \eeq
where Krylov wave function is $\phi_n(t) = i^{-n} \big\la \Psi_n \big| \Psi_0(t) \big\ra$. 

So, dynamical properties of a quantum system, such as chaos and thermalization, are closely related to properties of the Krylov wave function. The most important properties are position of the wavepacket $\phi_n(t)$, which is determined by the Krylov complexity (K-complexity):
\beq K(t) = \sum_{n=0}^{D-1} n \left| \phi_n(t) \right|^2, \eeq
and its localization length, which is captured by the exponent of the Krylov entropy (K-entropy):
\beq S(t) = - \sum_{n=0}^{D-1} \left| \phi_n(t) \right|^2 \log \left| \phi_n(t) \right|^2. \eeq
In a chaotic system, $K(t)$ grows exponentially and eventually saturates at $\overline{K(t)} \sim d$, whereas Krylov wave function quickly becomes \textit{delocalized}. In an integrable system, $K(t)$ grows slower and saturates at a much smaller value, $\overline{K(t)} \ll d$, since Krylov wave function is \textit{localized} and state $\left| \psi (t) \right\ra$ explores only a small fraction of the Hilbert space. So, Krylov delocalization serves as a signature of quantum chaos in continuous time dynamics~\cite{Gorsky:2019,Trigueros:2021,Rabinovici:2021,Menzler:2024,Alaoui:2023,Alishahiha:2024}.

\textit{\textbf{Szeg\H{o} algorithm---}} Now, let us generalize construction~\eqref{eq:Lanczos} to Floquet case, i.e., to a periodic Hamiltonian $H(t + T) = H(t)$. As long as we restrict the evolution of a Floquet system to discrete times $t = n T$, the Krylov subspace of a seeding state $| \psi \ra$ has the following form:
\beq \label{eq:subspace-U}
| \psi(n T) \ra \in \mathrm{span}\!\left\{ | \psi \ra, U | \psi \ra, U^2 | \psi \ra, \cdots, U^{D-1} | \psi \ra \right\}, \eeq
where we introduce the evolution operator in discrete time $U =  \mathrm{Texp}\!\left[- i \int_0^T dt \, H(t)\right]$. 

To perform the generalization, we recall that Krylov construction~\eqref{eq:Lanczos} is closely related to the theory of orthogonal polynomials~\cite{Ismail:2005,Ismail:2020,Muck:2022}. Indeed, we can map $| \psi \ra \to 1$, $H \to x$, and $| O_n \ra = \mathcal{P}_n(H) | \psi \ra \to \mathcal{P}_n(x)$, where $\{ \mathcal{P}_n(x) \}$ is a sequence of orthonormal polynomials on the real line. Such a sequence is fixed by recurrence relations~\eqref{eq:Lanczos}. Similarly, we can map $| \psi \ra \to 1$, $U \to z$, $\mathcal{Q}_n(U) | \psi \ra \to \mathcal{Q}_n(z)$, where $\{\mathcal{Q}_n(z)\}$ is a sequence of orthonormal polynomials on the unit circle $|z| = 1$. This sequence is fixed by Szeg\H{o} recurrence relations~\cite{Szego,Simon,Ismail:2005,Ismail:2020}. Hence, we can apply the Szeg\H{o} algorithm to construct a complete orthonormal basis $\left\{ | \Phi_n \ra = \mathcal{Q}_n(U) | \psi \ra \right\}$ in the Krylov subspace~\eqref{eq:subspace-U}:
\beq \label{eq:Szego} 
\begin{gathered}
    \bar{\alpha}_n = \la \tilde{\Phi}_n | U | \Phi_n \ra, \qquad \rho_n^2 = 1 - |\alpha_n|^2, \\
    \bem | \Phi_{n+1} \ra \\ | \tilde{\Phi}_{n+1} \ra \eem = \frac{1}{\rho_n} \bem U & -\bar{\alpha}_n \\ -\alpha_n \, U & 1 \eem \bem | \Phi_n \ra \\ | \tilde{\Phi}_n \ra \eem,
\end{gathered} \eeq
where $| \Phi_0 \ra = | \tilde{\Phi}_0 \ra = | \psi \ra / \| \psi \|$ and $\rho_{D-1} = 0$ \footnote{Note that the set of states $| \tilde{\Phi}_n \ra$ is normalized but not orthogonal: $\la \tilde{\Phi}_n | \tilde{\Phi}_n \ra = 1$ but $\la \tilde{\Phi}_m | \tilde{\Phi}_n \ra \neq 0$ for $m \neq n$. However, these states are orthogonal to Krylov basis states generated after step $n$: $\la \Phi_m | \tilde{\Phi}_n \ra = 0$ for $m > n$. Besides, $ \la \tilde{\Phi}_m | U | \tilde{\Phi}_n \ra = 0$ and $ \la \tilde{\Phi}_m | U | \Phi_n \ra = 0$ for $m > n$.}. Complex numbers $\alpha_n$, $| \alpha_n| \le 1$, are called the Verblunsky coefficients. We emphasize that computational complexity of this algorithm is equal to complexity of the Lanczos algorithm and significantly smaller than complexity of Arnoldi iteration or direct diagonalization of~$U$.

The unitary evolution operator $U$ in the basis $\{ | \Phi_n \ra \}$ has an upper Hessenberg form:
\beq \label{eq:Hessenberg}
\la \Phi_m | U | \Phi_n \ra = \begin{cases}
    -\bar{\alpha}_n \alpha_{m-1} \prod_{k=m}^{n-1} \rho_k, \; & m < n + 1, \\
    \rho_n, \; & m = n + 1, \\
    0 \; & m > n + 1,
\end{cases} \eeq
where we introduce $\alpha_{-1} = -1$ for convenience. Note that Arnoldi iteration~\cite{Yates:2021a,Qi:2024,Nizami:2023,Nizami:2024,Suchsland:2023,Scialchi:2024} applied to unitary $U$ reproduces matrix~\eqref{eq:Hessenberg}, but cannot discern its underlying structure. We also emphasize that unitary operator~\eqref{eq:Hessenberg}
describes a \textit{non-local} evolution on the Krylov chain.

\begin{figure}
    \centering
    \includegraphics[width=\linewidth]{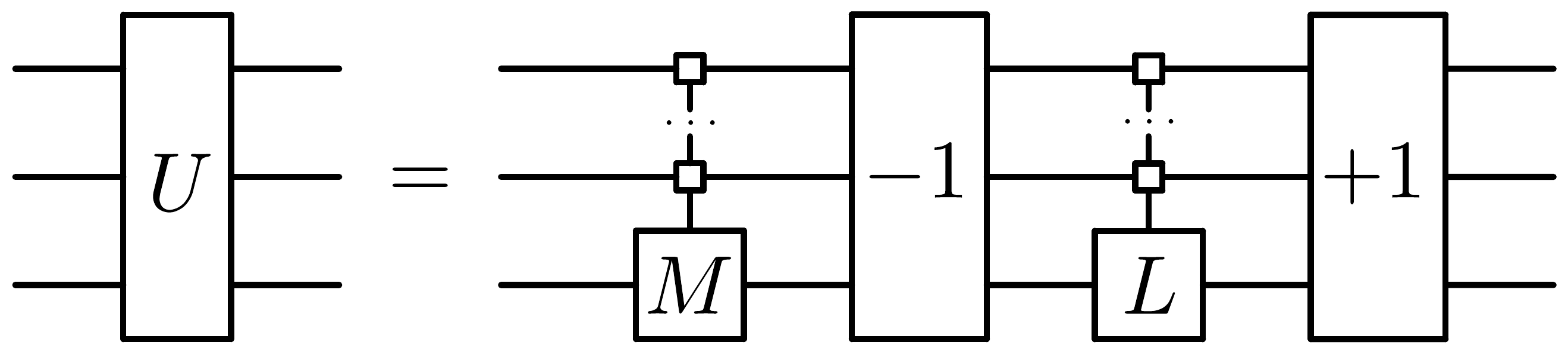}
    \caption{Quantum circuit implementing an arbitrary unitary $U$ in the CMV basis~\eqref{eq:CMV}. The seeding state $| \psi \ra = | 0 0 \cdots 0 \ra$.}
    \label{fig:circuit}
\end{figure}

\textit{\textbf{CMV basis---}}  Szeg\H{o} algorithm can be optimized to produce a five-diagonal matrix representation of~$U$. To show this, we define another complete and orthonormal basis, the Cantero---Moral---Velazquez (CMV) basis~\cite{CMV}:
\beq | P_{2n} \ra = U^{-n} | \tilde{\Phi}_{2n} \ra, \quad | P_{2n+1} \ra = U^{-n} | \Phi_{2n+1} \ra. \eeq
The algorithm that builds the CMV basis from a scratch has the same complexity as the Lanczos algorithm:
\beq \label{eq:CMV} \begin{aligned}
    | P_{2n+1} \ra &= \frac{\rho_{2n-1}}{\rho_{2n}} U | P_{2n-1} \ra - \! \left[ \frac{\bar{\alpha}_{2n-1}}{\rho_{2n}} U  + \frac{\bar{\alpha}_{2n}}{\rho_{2n}} \right] \! | P_{2n} \ra, \\
    \bar{\alpha}_{2n+1} &= \left\la P_{2n} | U | P_{2n+1} \right\ra / \rho_{2n}, \\
    | P_{2n+2} \ra &= \frac{\rho_{2n}}{\rho_{2n+1}} U^\dag | P_{2n} \ra - \! \left[ \frac{\alpha_{2n}}{\rho_{2n+1}} U^\dag  + \frac{\alpha_{2n+1}}{\rho_{2n+1}} \right] \! | P_{2n+1} \ra, \\
    \bar{\alpha}_{2n+2} &= \left\la P_{2n+2} | U | P_{2n+1} \right\ra / \rho_{2n+1},
\end{aligned} \eeq
with initial conditions $| P_0 \ra = | \psi \ra / \| \psi \|$ and $\bar{\alpha}_{-1} = -1$ and $\bar{\alpha}_0 = \la P_0 | U | P_0 \ra$ until $\rho_{D-1} = 0$. However, the choice of the CMV basis puts the evolution operator $U$ to a computationally convenient five-diagonal form:
\beq \label{eq:five-diagonal} \begin{gathered}
    \la P_m | U | P_n \ra = \sum_{k=0}^{D-1} L_{mk} M_{kn}, \\
    L = 1 \oplus \Theta_1 \oplus \Theta_3 \oplus \cdots, \quad M = \Theta_0 \oplus \Theta_2 \oplus \cdots,
\end{gathered} \eeq
where $D \times D$ matrices $L$ and $M$ consist of $2 \times 2$ unitary blocks and $1 \times 1$ identity blocks at the beginning and end:
\beq \Theta_n = \bem \bar{\alpha}_n & \rho_n \\ -\rho_n & \alpha_n \eem = \bem e^{-i \chi_n} \cos \theta_n & \sin \theta_n \\ -\sin \theta_n & e^{i \chi_n} \cos \theta_n \eem. \eeq
If we assume that the CMV basis coincides with the computational basis of an $L$-qubit quantum circuit ($D \le 2^L$), evolution operator~\eqref{eq:five-diagonal} is implemented using two quantum multiplexors and adding operations, see Fig.~\ref{fig:circuit}~\footnote{To restrict the dimension of the subspace simulated on the $L$-qubit circuit to $D < 2^L$, one simply needs to set $|\alpha_n| = 1$ for $n \ge D$, which decouples this subspace from the rest of the Hilbert space.}. Hence, an \textit{arbitrary} quantum dynamics in a Krylov basis is simulated using $\sim 2^L$ CNOT gates. Of course, this result does not contradict the theoretical bound~\cite{Shende:2004,Shende:2006} because Krylov basis depends on the seeding state $| \psi \ra$. 

Representation~\eqref{eq:five-diagonal} also implies that we can define $\varphi_n(t) = \big\la P_n \big| P_0(t) \big\ra$ and map the dynamics of a seeding state $| \psi \ra$ into a one-dimensional tight-binding model:
\beq \label{eq:tight-binding} \begin{aligned}
    \pd_t^2 \varphi_n(t) = &\left( - \alpha_n \bar{\alpha}_{n-1} - \bar{\alpha}_n \alpha_{n-1} - 2 \right) \varphi_n(t) \\ 
    &+ 2 \left( \lambda_{n+1} \pd_n \bar{\alpha}_{n-1} + \lambda_n \pd_n \alpha_{n-1} \right) \rho_{n-1} \, \varphi_{n-1}(t) \\ 
    &+ 2 \left( \lambda_{n+1} \pd_n \bar{\alpha}_n + \lambda_n \pd_n \alpha_n \right) \rho_n \, \varphi_{n+1}(t) \\ 
    &+ \rho_{n-1} \rho_{n-2} \, \varphi_{n-2}(t) + \rho_n \rho_{n+1} \, \varphi_{n+2}(t),
\end{aligned} \eeq
where $\lambda_n = \left[ 1 - (-1)^n \right]/2$ and initial conditions are
\beq \label{eq:CMV-ic}
\varphi_n(0) = \delta_{n,0} \quad \mathrm{and} \quad \varphi_n(T) = \bar{\alpha}_0 \delta_{n,0} + \rho_0 \delta_{n,1}. \eeq
For convenience, we introduce discrete derivatives over the chain index, $\pd_n \alpha_n = (\alpha_{n+1} - \alpha_{n-1})/2$, and time, $\pd_t^2 \varphi_n(t) = \varphi_n(t+T) + \varphi_n(t - T) - 2 \varphi_n(t)$.

In the high-frequency limit $T \to 0$, coefficients $\alpha_n$ are expressed in terms of Lanczos coefficients, so bases~\eqref{eq:Szego} and~\eqref{eq:CMV} reduce to the standard Krylov basis~\eqref{eq:Lanczos}, see Appendix~\ref{sec:Szego-to-Lanczos}. In this limit, even and odd sites of the chain are decoupled, so model~\eqref{eq:tight-binding} reproduces~\eqref{eq:tight-binding-Lanczos}.

We also emphasize that the tight-binding model~\eqref{eq:tight-binding} reproduces the dynamics of the boundary operator in a non-interacting XY model, see Appendix~\ref{sec:XY}. So, model~\eqref{eq:tight-binding} generalizes the approach of~\cite{Yeh:2023,Yeh:2024} from vectorized operators to arbitrary state dynamics.

\textit{\textbf{Signatures of chaos and integrability---}} Inspired by eigenvalue properties of matrix ensembles~\cite{Killip:2006,Killip:2004,Dumitriu:2002,Das:2021,Balasubramanian:2022-2}, we conjecture that the sequence of Verblunsky coefficients generated by a typical seeding state in a quantum Floquet system falls into one of the following classes~\footnote{For a finite $d$, Verblunsky coefficients might deviate from these distributions, but the average magnitude of deviations goes to zero as $d \to \infty$}:
\begin{enumerate}
    \item[(\textit{i})] \textit{Degenerate case}: $|\alpha_n|^2 < A/(d-n)^{1+\epsilon}$,
    \item[(\textit{ii})] \textit{Chaotic case}: $|\alpha_n|^2 \sim A / \left[ \beta (d-n) / 2 + 1 \right]$,
    \item[(\textit{iii})] \textit{Integrable case}: $|\alpha_n|^2 > A/(d-n)^{1-\epsilon}$,
\end{enumerate}
where $\epsilon > 0$, $\beta > 0$, constant $A > 0$ is chosen such that all coefficients lie inside the unit circle, and $1 \ll n < d$.

To illustrate this classification, we consider \textit{random} Verblunsky coefficients independently distributed on the unit disk with the following probability density functions:\begin{subequations} \label{eq:ensemble} \begin{align}
p_n^{(i)}(\alpha_n) &\sim \left(1 - |\alpha_n|^2 \right)^{[ (d - n)^{(1 + \epsilon)} - 2 ]}, \\
p_n^{(ii)}(\alpha_n) &\sim \left(1 - |\alpha_n|^2 \right)^{[ \beta (d - n)/2 - 1 ]}, \\
p_n^{(iii)}(\alpha_n) &\sim \left(1 - |\alpha_n|^2 \right)^{[ (d - n)^{(1 - \epsilon)} - 2 ]}.
\end{align} \end{subequations}
First of all, we discuss the eigenvalue distribution of CMV matrices constructed using random Verblunsky coefficients~\footnote{Note that the eigenvalue distribution of the Krylov chain is only an indirect probe of chaoticity of the original many-body dynamics. The main loophole is that the dimension of the Krylov subspace might be smaller than the total Hilbert space dimension, so it captures only a fraction of energy levels. However, this is not the case for the examples we consider in this Letter.}. It was shown~\cite{Killip:2006} that these distributions converge to the clock ensemble~(\textit{i}), circular $\beta$-ensemble~(\textit{ii}), or Poisson ensemble~(\textit{iii}), respectively~\footnote{Circular $\beta$-ensemble includes the orthogonal ($\beta = 1$), unitary ($\beta = 2$), and symplectic ($\beta = 4$) ensembles.}. The mean level spacing ratio~\cite{Oganesyan:2007,Atas:2013} changes in the range $0.386 < \la r \ra < 1$ and grows with~$\beta$ in ensemble~(\textit{ii}), whereas ensembles~(\textit{i}) and~(\textit{iii}) saturate the upper and lower bounds of this inequality. In other words, energy levels are independently distributed in integrable ensemble~(\textit{iii}) and repel each other in chaotic ensemble~(\textit{ii}), which reproduces the conventional classification of ergodic systems~\cite{Haake,Stockmann,Berry:1977a,Bohigas:1984}. In 
ensemble~(\textit{i}), energy level repulsion is so strong that it becomes integrable but preserves some features of chaotic systems.

\begin{figure}
    \centering
    \includegraphics[width=\linewidth]{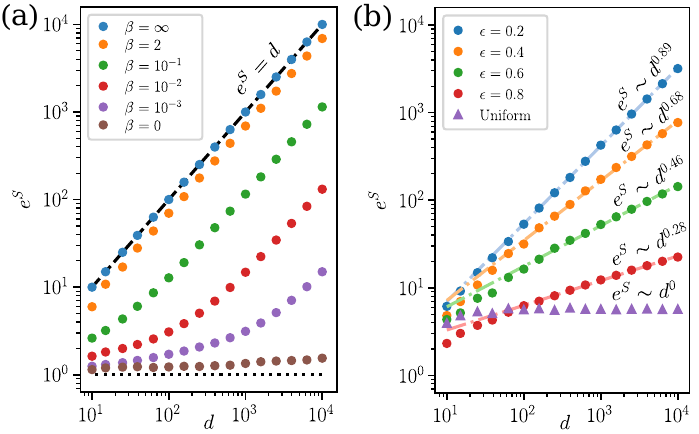}
    \caption{
    Krylov localization length averaged over time, initial position, and ensemble realization for different chaotic (a) and integrable (b) ensembles~\eqref{eq:ensemble}. In all cases, $K \sim e^S$.}
    \label{fig:ensemble-average}
\end{figure}

Furthermore, classification (\textit{i})--(\textit{iii}) reflects the dynamical properties of chaotic and integrable ensembles, i.e., implies Krylov localization in the integrable case and delocalization in the chaotic and degenerate cases. Let us show this using the same example with random Verblunsky coefficients~\eqref{eq:ensemble}. To that end, we write down the eigenvalue equation in the Krylov or CMV basis, average it over the corresponding ensembles, and estimate the probability distribution of an eigenvector $\varphi_n(\omega)$~\footnote{
In fact, eigenvectors of a non-averaged $U$ are localized around different sites and usually have different localization lengths. However, the qualitative prediction $e^S \sim d^{1 - \epsilon}$ re-emerges after averaging over eigenvectors.}:
\beq \frac{|\varphi_n(\omega)|^2}{|\varphi_0(\omega)|^2} \sim \prod_{k=0}^{n-1} \left\la \rho_k^2 \right\ra \sim \begin{cases} 1, \; &(i), \\ (1 - n/d)^{- 1 / 2 \beta}, \; &(ii), \\ \exp\!\left( - n / d^{1 - \epsilon} \right), \; &(iii). \end{cases} \eeq
So, we expect that Krylov wave function is delocalized in ensembles (\textit{i}) and (\textit{ii}), $e^S \sim d$, but localized in integrable ensembles (\textit{iii}), $e^S \sim d^{1 - \epsilon} \ll d$ (cf. Sec.~12.7 of~\cite{Simon}). Such a localization was also proved rigorously for selected distributions, e.g., see~\cite{Teplyaev:1991,Cedzich:2019,Zhu:2021}. Numerical calculations confirm this qualitative analysis, see Fig.~\ref{fig:ensemble-average}.

\begin{figure}[b]
    \centering
    \includegraphics[width=\linewidth]{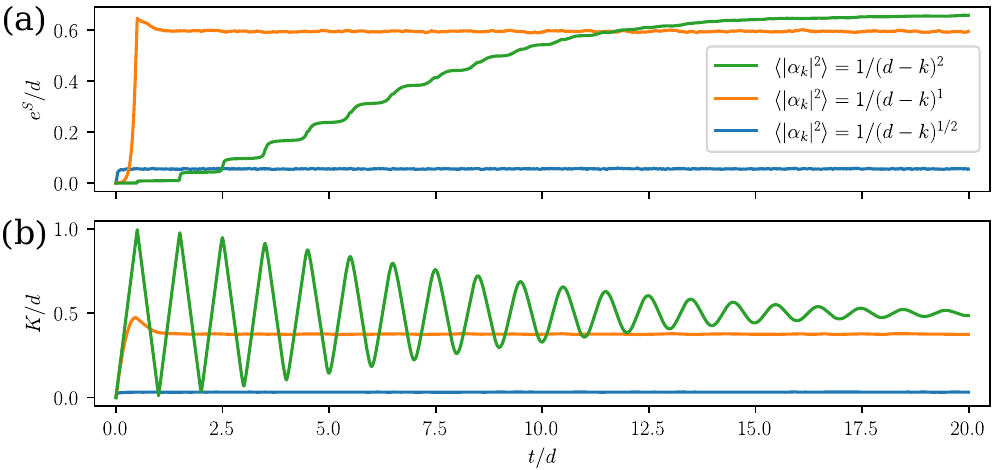}
    \caption{
    Ensemble-averaged localization length (a) and position (b) of an initially localized wave packet propagating along the chain with random coefficients~\eqref{eq:ensemble} and length $d = 2000$.}
    \label{fig:ensemble-time}
\end{figure}

Another way to see localization-delocalization transition is to notice that the tight-binding model~\eqref{eq:tight-binding} reduces to a massless Klein-Gordon equation when $|\alpha_n|^2 \ll 1$:
\beq \label{eq:KG}
\pd_t^2 \varphi_n(t) \approx 4 \pd_n \pd_n \varphi_n(t) + \mO( \alpha_n^2), \eeq
where $\mO( \alpha_n^2)$ is a small noise term. Solutions to an unperturbed Eq.~\eqref{eq:KG} with initially localized Krylov wave function $\varphi_n(0) = \delta_{n,0}$ are given by a sum of localized wave packets. In a degenerate ensemble~(\textit{i}), wave packets remain localized until they reach the far end of the Krylov chain. So, they explore the whole Hilbert space, but stay localized up to $t \gg d$. In a beta ensemble~(\textit{ii}), the noise term smoothes the propagating wave packet, so its localization length reaches $e^S \sim d$ as soon as its center approaches the far half of the Krylov chain. In a Poisson ensemble~$(iii)$, the noise term is so strong that the wave packet cannot propagate at all, see Fig.~\ref{fig:ensemble-time}.

So, the behavior of Verblunsky coefficients serves as a signature of quantum chaos and integrability similarly to behavior of Lanczos coefficients~\cite{Parker:2018}. In particular, in non-degenerate quantum systems with $\la r \ra < 1$, \textit{the slowest possible growth of Verblunsky coefficients is achieved for chaotic systems}. In what follows, we will give several examples where such kinds of behavior indeed emerge.

\textit{\textbf{Semiclassical Krylov localization---}} We start with the \textit{kicked top model}, a paradigmatic example of a quantum Floquet system~\cite{Haake,Stockmann,Haake:1987,Izrailev:1990,Rozenbaum:2016,Vallini:2024} with remarkable experimental realizations~\cite{Chaudhury:2009,Neill:2016}:
\beq \label{eq:kicked-top}
U_\mathrm{top} = \exp\!\left( -i \frac{\kappa_x}{2 J} J_x^2 \right) \exp\!\left(-i \frac{\kappa_z}{2 J} J_z^2 \right) \exp\!\left[ -i \left( \mathbf{b} \cdot \mathbf{J} \right) \right]. \eeq
Here, $\mathbf{J} = (J_x, J_y, J_z)$, $[ J_k, J_l] = i \epsilon_{klm} J_m$ is the angular momentum operator. The square $\mathbf{J}^2 = J(J+1)$ is conserved, so we restrict our discussion to the $d = 2J+1$ dimensional Hilbert space. We also emphasize that $U_\mathrm{top}$ has reflection symmetry, so we further restrict its dynamics to one of the invariant sectors. 

In the semiclassical limit $\hbar = 1/J \to 0$, Heisenberg evolution~\eqref{eq:kicked-top} reproduces the dynamics of a classical unit spin $\mathbf{j} = \mathbf{J}/J$. with Poisson bracket $\{ j_k, j_l \} = \epsilon_{klm} j_m$. This dynamics can be regular, chaotic, or mixed depending on the parameters of the model. We set $\kappa_z = 0$, $\mathbf{b} = (1.7,0,0)$ to study regular (integrable) regime and $\kappa_z = 0.5$, $\mathbf{b} = (0,1.7,0)$ to study chaotic regime~\cite{Haake:1987,Izrailev:1990}. In~both cases, we generate an ensemble of evolution operators~\eqref{eq:kicked-top} by varying one of the angular momenta in the range $10 \le \kappa_x \le 12$ with the step $\delta \kappa_x = 0.05$~\footnote{This approach allows us to get cleaner results for small Hilbert space dimensions, but superfluous for $d > 10^3$.}. We take a spin coherent state $| \Omega \ra$ as a natural seeding state for the Krylov construction.

\begin{figure}
    \centering
    \includegraphics[width=\linewidth]{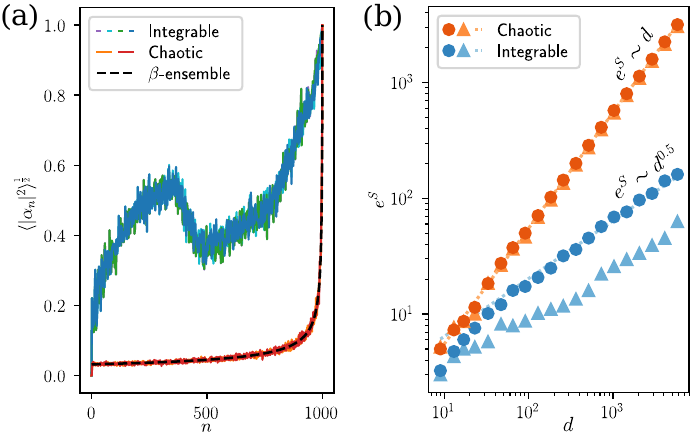}
    \caption{(a) Distribution of Verblunsky coefficients in the integrable and chaotic kicked top ($J = 1000$). (b) Localization length of a seeding coherent state (circles) and average localization length of all Krylov basis elements (triangles).}
    \label{fig:top}
\end{figure}

In the regular regime, classical phase space breaks into periodic trajectories. So, any initially localized distribution evolves in a region of area~$A$ sandwiched between two adjacent trajectories. Semiclassical quantization maps periodic trajectories to eigenstates of $U_\mathrm{top}$, each of which is concentrated in a phase space region of area $A_0 \sim \hbar$~\cite{Berry:1977b}. A coherent state $| \Omega \ra$ sweeps the area $A \sim L \sqrt{\hbar}$ when $\Omega$ belongs to a classical trajectory of finite length $L$, so it spreads over $D \sim L / \sqrt{\hbar}$ eigenstates. Therefore, K-complexity and K-entropy of such a state are proportional to $K \sim e^S \sim \sqrt{d} \ll d$,~while Verblunsky coefficients follow an integrable distribution~(\textit{iii}). Numerical simulations support this reasoning (Fig.~\ref{fig:top}).

In the chaotic regime, classical evolution brings any point arbitrarily close to any other point. So, an initially localized classical distribution uniformly spreads over the phase space, while the corresponding quantum state uniformly covers the Hilbert space~\cite{Berry:1977b}. Conversely, we can switch to a phase-space-local basis (e.g., basis of $J_z$ eigenstates) and think of $U_\mathrm{top}$ as a random unitary transformation of complex vectors written in this basis. Applying Szeg\H{o} algorithm to such an evolution, we obtain that Verblunsky coefficients follow the chaotic distribution~(\textit{ii}) with~$\beta = 2$. This implies delocalization of the Krylov wave function, $K \sim e^S \sim d$, see Fig.~\ref{fig:top}.

\textit{\textbf{Spin chains---}} To show that similar distributions of Verblunsky coefficients emerge far from the semiclassical limit, we consider the \textit{kicked Ising chain}~\cite{Prozen:2000,Prozen:2002,Prozen:2007,Prozen:2018,Prozen:2019,Akila:2016}:
\beq \label{eq:Ising}
U_\mathrm{KI} = \exp\!\left( -i J \sum_{k=0}^{L-1} \sigma_k^z \sigma_{k+1}^z \right) \exp\!\left[ -i \sum_{k=0}^{L-1} \left( \mathbf{b} \cdot \boldsymbol{\sigma}_k \right) \right], \eeq
with $\sigma^{x,y,z}_k$ being the Pauli matrices on site $k$. We assume periodic boundary conditions, $\mathbf{\sigma}_L = \mathbf{\sigma}_0$, and restrict the dynamics to one of translation-invariant sectors.

To generate different ensembles of $U_\mathrm{KI}$, we fix the exchange constant $J = 0.7$ and vary the magnitude of the magnetic field in the range $1.13 \le |\textbf{b}| \le 1.41$ with the step $\delta b = 0.007$. We fix the direction $\mathbf{b}/|\textbf{b}| = (1,0,0)$ to study integrable behavior and $\mathbf{b}/|\textbf{b}| = (1/\sqrt{2},0,1/\sqrt{2})$ to study chaotic behavior~\cite{Prozen:2000,Prozen:2002}. We choose a random superposition of several computational basis states as a natural low-entanglement seeding state.

\begin{figure}
    \centering
    \includegraphics[width=\linewidth]{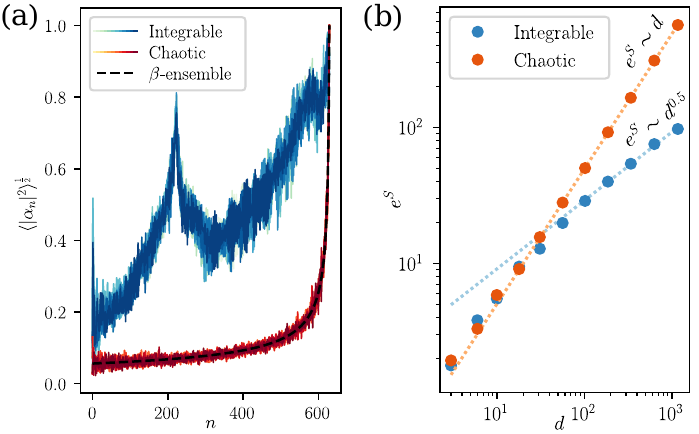}
    \caption{(a) Distribution of Verblunsky coefficients in the integrable and chaotic kicked Ising chain ($L = 13$). (b) Localization length of a seeding state averaged over random seeding states from different symmetry sectors.}
    \label{fig:chain}
\end{figure}

Similarly to the kicked top, kicked Ising chain produces Verblunsky coefficients with distribution (\textit{iii}) in the integrable case and distribution (\textit{ii}) in the chaotic case. So, transition from chaos to integrability is again marked by Krylov localization of low-entanglement states, see Fig.~\ref{fig:chain}. In other words, integrable evolution drags such states through small periodic cycles, whereas chaotic evolution explores the Hilbert space uniformly.

\textit{\textbf{Convergence to maximally ergodic regime---}} It was recently conjectured that Krylov basis of a generic chaotic Floquet system converges to ``maximally ergodic'' states with vanishing autocorrelation functions~\cite{Suchsland:2023,Scialchi:2024,Yeh:2023,Yeh:2024}. Our observations complement this conjecture and emphasize that the thermodynamic limit should be taken with caution. Indeed, the conjectured convergence to ``maximally ergodic'' basis implies that \textit{all} Verblunsky coefficients approach $|\alpha_n|^2 \to 0$ for $n \gg 1$, i.e., belong to case~(\textit{i}) from our classification. This, in turn, automatically implies that the spectrum of evolution operator approaches the clock ensemble with $\la r \ra \approx 1$, which is almost never true for a generic chaotic system.

In fact, Verblunsky coefficients of chaotic system are small but non-zero: $|\alpha_n|^2 \sim 1/d$ for $1 \ll n \ll d$, where~$d$ is the dimension of the Hilbert space. In the limit $d \to \infty$, infinitely many coefficients go to zero, which corroborates the ``maximally ergodic'' conjecture; however, these states span a vanishingly small fraction of the   full Hilbert space. This difference seems innocuous, but it substantially affects state dynamics for large but finite $d$. In particular, it accelerates the spread of initially localized distribution in the Krylov space (Fig.~\ref{fig:ensemble-time}). So, we suggest to use asymptotic distribution~(\textit{ii}) instead of maximally ergodic bath approximation~(\textit{i}) to improve the accuracy of numerical simulations in chaotic Floquet systems.

\begin{figure}[t]
    \centering
    \includegraphics[width=\linewidth]{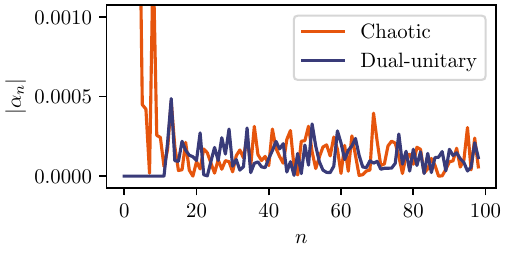}
    \caption{Verblunsky coefficients of a local operator in dual-unitary and general chaotic Ising model~\eqref{eq:Ising} with $L=14$.}
    \label{fig:dual-unitary}
\end{figure}

We emphasize that the early-time behavior of Verblunsky coefficients is still determined by the features of the system and initial state. For example, local Pauli operators in a dual-unitary circuit on $L$ sites generate exactly zero Verblunsky coefficients up to $n \sim L$, but reproduce general chaotic evolution with $|\alpha_n|^2 \sim 1/d \sim 1/ 4^L$ at larger times $L \ll n \ll d$, see Fig.~\ref{fig:dual-unitary}. 

\textit{\textbf{Conclusion---}} We suggested a new approach to Krylov basis construction in quantum Floquet systems. Our algorithm works substantially faster than other existing approaches~\cite{Nizami:2023,Nizami:2024,Suchsland:2023,Scialchi:2024,Yeh:2023,Yeh:2024} and brings the unitary evolution operator to a five-diagonal form, which is extremely convenient for simulations on both classical and quantum computers. So, our algorithm streamlines numerical experiments and allows them to tackle larger Hilbert space dimensions for a given quantum system. Furthermore, it provides a toolkit for engineering quantum Floquet systems with desired quantum dynamics, which is especially useful for simulations on small quantum computers. 

We emphasize that for the sake of generality, we mainly focused on the state evolution. Note, however, that we can always map an operator $ O = \sum_{i,j=1}^d O_{ij} |\psi_i \ra \la \psi_j|$ to a state $| O \ra = \sum_{i,j=1}^d O_{ij} |\psi_i \ra \otimes | \psi_j \ra$, i.e., to an element of $d^2$-dimensional vector space endowed with an inner product $(A | B) = \tr\big(A^\dag B\big)/d$. So, our construction works equally well for the operator dynamics. The only difference is that the operator evolution is described by the superoperator $\mU: \mU O = U^\dag O U$, which has some additional structure compared to a simple $U$. This additional structure confines all Verblunsky coefficients to a real segment $I = [-1, 1]$ instead of the unit disk, which implies several profound relations, see Appendix~\ref{sec:Szego-to-Lanczos}.

We also suggested a classification of ergodic Floquet systems based on the asymptotic behavior of Verblunsky coefficients. This classification reproduces the conventional classification based on energy level repulsion, but provides additional insights into state dynamics. To illustrate our conjecture, we obtained the proposed distributions of Verblunsky coefficients for simple seeding states in random matrix ensembles, kicked top, and kicked Ising chain. We emphasize that in general, distribution of Verblunsky coefficients and Krylov localization length depends on the initial state~\footnote{The simplest example is quantized system with a mixed phase space, where regular and chaotic regions coexist.
}; the only exceptions are chaotic circular ensembles with $\beta = 1,2,4$, which generate distribution (\textit{ii}) for an arbitrary initial state. We will study this dependence in the future work.

There are many other directions for future research. First, we expect that the behavior of Veblunsky coefficients might indicate the emergence of quantum scars and Hilbert space fragmentation~\cite{Serbyn:2020,Chandran:2022,Moudgalya:2021,Moudgalya:2021,Sala:2020,Khemani:2019,Motrunich:2021}. Second, our approach provides an alternative perspective for thermalization and information spreading in random and dual-unitary quantum circuits~\cite{Fisher:2022,Khemani:2017,Arute:2019,Bertini:2019,Fritzsch:2021,Gopalakrishnan:2019,Lerose:2020}. Finally, our numerical algorithm can be applied for simulations of such remarkable driven systems as Floquet topological insulators~\cite{Po:2016,Nathan:2019} and discrete time crystals~\cite{Wilczek:2012,Khemani:2016,Else:2016,Choi:2017,Zhang:2017}.

\textit{\textbf{Acknowledgments---}} We thank David A. Huse, Alexander Gorsky, Anatoly Dymarsky, Dmitry Abanin, Philippe Suchsland, and Elizaveta Trunina for valuable discussions.

\appendix

\section{Reduction from Szeg\"{o} to Lanczos}
\label{sec:Szego-to-Lanczos}

Let us show that Krylov bases~\eqref{eq:Szego} and~\eqref{eq:CMV} reduce to the standard Krylov basis~\eqref{eq:Lanczos} in the limit $T \to 0$. Exploiting the theory of orthogonal polynomials, we rewrite the $n$-th element of basis~\eqref{eq:Szego} in the determinant form:
\beq | \Phi_n \ra = \frac{1}{\sqrt{\Delta_n \Delta_{n-1}}} \det \bem c_0 & c_1 & \cdots & c_n \\ c_{-1} & c_0 & \cdots & c_{n-1} \\
\cdots & \cdots & \ddots & \cdots \\ c_{1-n} & c_{2-n} & \cdots & c_1 \\ | \psi \ra & U | \psi \ra & \cdots & U^n | \psi \ra \eem, \eeq
where we introduce the Toeplitz matrix:
\beq \Delta_n = \det \bem c_0 & c_1 & \cdots & c_n \\ c_{-1} & c_0 & \cdots & c_{n-1} \\
\cdots & \cdots & \ddots & \cdots \\ c_{-n} & c_{1-n} & \cdots & c_0 \eem, \eeq
and coefficients $c_n = \la \psi | U^n | \psi \ra$. Elements of the standard basis~\eqref{eq:Lanczos} are represented in a similar determinant form:
\beq | \Psi_n \ra = \frac{1}{\sqrt{D_n D_{n-1}}} \det \bem \mu_0 & \mu_1 & \cdots & \mu_n \\ \mu_1 & \mu_2 & \cdots & \mu_{n+1} \\
\cdots & \cdots & \ddots & \cdots \\ \mu_{n-1} & \mu_{n-2} & \cdots & \mu_{2n-1} \\ | \psi \ra & H | \psi \ra & \cdots & H^n | \psi \ra \eem, \eeq
with the Hankel matrix:
\beq D_n = \det \bem \mu_0 & \mu_1 & \cdots & \mu_n \\ \mu_1 & \mu_2 & \cdots & \mu_{n+1} \\
\cdots & \cdots & \ddots & \cdots \\ \mu_n & \mu_{n+1} & \cdots & \mu_{2n} \eem, \eeq
and coefficients $\mu_n = \la \psi | H^n | O \ra$. In the limit $T \to 0$, we can approximate $U = \exp(-i T H)$, where the instantaneous Hamiltonian is $H = i (\pd U / \pd T)_{T=0}$. Hence, the leading approximation to coefficients~$c_n$ is directly expressed through $\mu_n$:
\beq c_n \approx \la \psi | e^{- i n T H} | \psi \ra = \sum_{k=0}^\infty \frac{(-i n T)^k}{k!} \mu_k. \eeq
In the same limit, Toeplitz matrices reduce to Hankel matrices:
\beq \Delta_n \approx T^{n^2 + n} D_n, \eeq
so Krylov bases~\eqref{eq:Szego} and~\eqref{eq:CMV} coincide with Krylov basis~\eqref{eq:Lanczos} up to phase and higher-order corrections in $T$:
\beq \label{eq:approximate-basis}
| P_n \ra \approx | \Phi_n \ra \approx (-i)^n | \Psi_n \ra. \eeq
Similarly, we prove that Verblunsky coefficients $\alpha_n$ have the following approximate expansion in terms of Lanczos coefficients:
\beq \alpha_n \approx (-1)^n \! \left[ 1 + i T A_n - \frac{T^2}{2} \! \left( b_{n+1}^2 - A_n^2 \right) + \cdots \right], \eeq
where we denote $A_n = \sum_{k=0}^n a_k$ for brevity. We also note that the leading approximation to $\rho_n$ does not depend on the diagonal Lanczos coefficients: $\rho_n \approx T \left| b_{n+1} \right| + \cdots$. 

Surprisingly, Lanczos algorithm determines the sequence $\alpha_n$ and matrix $U$ \textit{exactly} even away from the limit $T \to 0$ when all coefficients $\alpha_n$ are \textit{real}. In this case, we can employ the construction suggested in~\cite{DVG,CMV:2020}. Namely, we define the Hamiltonian $\mH$:
\beq \mH = i \left( U^{1/2} - U^{-1/2} \right), \eeq
redefine the inner product and norm:
\beq ( A | B ) \to \left( A | (1 - \mH / 2) | B \right), \eeq
and generate sequences $\{ a_n \}$ and $\{ b_n \}$ using the standard Lanczos algorithm~\eqref{eq:Lanczos}. These sequences are unambiguously related to coefficients $\alpha_n$ of the original CMV basis:
\beq a_n = (-1)^n \left( \alpha_n + \alpha_{n-1} \right), \qquad b_n = \rho_{n-1}. \eeq
Hence, we easily restore coefficients $\alpha_n$ and matrix $U$:
\beq \label{eq:Lanczos-to-Szego}
\alpha_n = (-1)^n \sum_{k=0}^n a_k + (-1)^n. \eeq
This construction is especially useful for a Hermitian seeding operator $O^\dag = O$, which naturally produces real coefficients $\alpha_n$. In particular, correspondence~\eqref{eq:Lanczos-to-Szego} allows us to use the most efficient algorithms, such as Partial Re-Orthogonalization~\cite{Rabinovici:2020}, and enlarge the accessible size of operator space by several orders of magnitude compared to straightforward Arnoldi iteration. The obvious disadvantage is the need to calculate $U^{1/2}$ first. To bypass this disadvantage, one may consider the evolution in doubled discrete time generated by $U^2$. Since chaotic systems remain chaotic and integrable systems remain integrable no matter what time scale we choose, we expect that the change $U \to U^2$ does not qualitatively affect operator dynamics and coefficients $\alpha_n$. In fact, we numerically confirmed for several models that K-complexity and K-entropy in CMV bases generated by $U$ and $U^2$ coincide up to insignificant $\mO(1)$ factors.

\section{Inhomogeneous XY model}
\label{sec:XY}

An arbitrary unitary dynamics in Krylov space~\eqref{eq:CMV} of dimension $d$ is mapped to an inhomogeneous XY model on $d$ sites:
\beq \label{eq:XY}
    \begin{aligned}
    U_\mathrm{XY} &= \prod_{k=1}^{d/2 - 1} U_{2k-1,2k} \prod_{k=0}^{d/2 - 1} U_{2k,2k+1}, \\
    U_{k,k+1} &= e^{\frac{i \chi_k}{4} (\sigma_k^z - \sigma_{k+1}^z)} e^{\frac{i \theta_k}{2} (\sigma_k^x \sigma_{k+1}^x + \sigma_k^y \sigma_{k+1}^y)} e^{\frac{i \chi_k}{4} (\sigma_k^z - \sigma_{k+1}^z)},
\end{aligned} \eeq
where $\chi_k \in [-\pi, \pi)$, $\theta_k \in [0, \pi/2]$ and $d$ is even. This model is effectively non-interacting. In fact, unitary~\eqref{eq:XY} is bilinear in terms of Jordan-Wigner creation and annihilation operators:
\beq c_k = \frac{1}{2} \left(\sigma_k^x - i \sigma_k^y \right) \prod_{j=0}^{k-1} \sigma_j^z, \qquad c_k^\dag = \frac{1}{2} \left(\sigma_k^x + i \sigma_k^y \right) \prod_{j=0}^{k-1} \sigma_j^z. \eeq
So, operator evolution~\eqref{eq:XY} preserves the subspace of one-particle fermion operators and reproduces the CMV matrix~\eqref{eq:five-diagonal} after identification $\alpha_k = e^{i \chi_k} \cos \theta_k$:
\beq U_\mathrm{XY}^\dag \left( \sum_{k=0}^d v_k c_k \right) U_\mathrm{XY} =\sum_{k=0}^d \left( LM v \right)_k c_k, \eeq
where $v_k$ are arbitrary complex numbers. In other words, this mapping identifies one-particle operators $c_k$ with the CMV basis vectors. In particular, boundary operator $c_0$ corresponds to the seeding state $| P_0 \ra = | \psi \ra / \| \psi \|$.

Finally, this model can be rewritten as an inhomogeneous transverse-field Ising model~\cite{Yeh:2023,Yeh:2024} when all Verblunsky coefficients are real, i.e., $\chi_k = 0$ for all $k$.
 
\bibliography{letter}

\begin{thebibliography}{117}%
\makeatletter
\providecommand \@ifxundefined [1]{%
 \@ifx{#1\undefined}
}%
\providecommand \@ifnum [1]{%
 \ifnum #1\expandafter \@firstoftwo
 \else \expandafter \@secondoftwo
 \fi
}%
\providecommand \@ifx [1]{%
 \ifx #1\expandafter \@firstoftwo
 \else \expandafter \@secondoftwo
 \fi
}%
\providecommand \natexlab [1]{#1}%
\providecommand \enquote  [1]{``#1''}%
\providecommand \bibnamefont  [1]{#1}%
\providecommand \bibfnamefont [1]{#1}%
\providecommand \citenamefont [1]{#1}%
\providecommand \href@noop [0]{\@secondoftwo}%
\providecommand \href [0]{\begingroup \@sanitize@url \@href}%
\providecommand \@href[1]{\@@startlink{#1}\@@href}%
\providecommand \@@href[1]{\endgroup#1\@@endlink}%
\providecommand \@sanitize@url [0]{\catcode `\\12\catcode `\$12\catcode `\&12\catcode `\#12\catcode `\^12\catcode `\_12\catcode `\%12\relax}%
\providecommand \@@startlink[1]{}%
\providecommand \@@endlink[0]{}%
\providecommand \url  [0]{\begingroup\@sanitize@url \@url }%
\providecommand \@url [1]{\endgroup\@href {#1}{\urlprefix }}%
\providecommand \urlprefix  [0]{URL }%
\providecommand \Eprint [0]{\href }%
\providecommand \doibase [0]{https://doi.org/}%
\providecommand \selectlanguage [0]{\@gobble}%
\providecommand \bibinfo  [0]{\@secondoftwo}%
\providecommand \bibfield  [0]{\@secondoftwo}%
\providecommand \translation [1]{[#1]}%
\providecommand \BibitemOpen [0]{}%
\providecommand \bibitemStop [0]{}%
\providecommand \bibitemNoStop [0]{.\EOS\space}%
\providecommand \EOS [0]{\spacefactor3000\relax}%
\providecommand \BibitemShut  [1]{\csname bibitem#1\endcsname}%
\let\auto@bib@innerbib\@empty
\bibitem [{\citenamefont {Deutsch}(1991)}]{Deutsch:1991}%
  \BibitemOpen
  \bibfield  {author} {\bibinfo {author} {\bibfnamefont {J.~M.}\ \bibnamefont {Deutsch}},\ }\bibfield  {title} {\bibinfo {title} {{Quantum statistical mechanics in a closed system}},\ }\href {https://doi.org/10.1103/PhysRevA.43.2046} {\bibfield  {journal} {\bibinfo  {journal} {Phys. Rev. A}\ }\textbf {\bibinfo {volume} {43}},\ \bibinfo {pages} {2046} (\bibinfo {year} {1991})}\BibitemShut {NoStop}%
\bibitem [{\citenamefont {Srednicki}(1994)}]{Srednicki:1994}%
  \BibitemOpen
  \bibfield  {author} {\bibinfo {author} {\bibfnamefont {M.}~\bibnamefont {Srednicki}},\ }\bibfield  {title} {\bibinfo {title} {{Chaos and quantum thermalization}},\ }\href {https://doi.org/10.1103/PhysRevE.50.888} {\bibfield  {journal} {\bibinfo  {journal} {Phys. Rev. E}\ }\textbf {\bibinfo {volume} {50}},\ \bibinfo {pages} {888} (\bibinfo {year} {1994})},\ \Eprint {https://arxiv.org/abs/cond-mat/9403051} {arXiv:cond-mat/9403051} \BibitemShut {NoStop}%
\bibitem [{\citenamefont {D'Alessio}\ \emph {et~al.}(2016)\citenamefont {D'Alessio}, \citenamefont {Kafri}, \citenamefont {Polkovnikov},\ and\ \citenamefont {Rigol}}]{DAlessio:2015}%
  \BibitemOpen
  \bibfield  {author} {\bibinfo {author} {\bibfnamefont {L.}~\bibnamefont {D'Alessio}}, \bibinfo {author} {\bibfnamefont {Y.}~\bibnamefont {Kafri}}, \bibinfo {author} {\bibfnamefont {A.}~\bibnamefont {Polkovnikov}},\ and\ \bibinfo {author} {\bibfnamefont {M.}~\bibnamefont {Rigol}},\ }\bibfield  {title} {\bibinfo {title} {{From quantum chaos and eigenstate thermalization to statistical mechanics and thermodynamics}},\ }\href {https://doi.org/10.1080/00018732.2016.1198134} {\bibfield  {journal} {\bibinfo  {journal} {Adv. Phys.}\ }\textbf {\bibinfo {volume} {65}},\ \bibinfo {pages} {239} (\bibinfo {year} {2016})},\ \Eprint {https://arxiv.org/abs/1509.06411} {arXiv:1509.06411} \BibitemShut {NoStop}%
\bibitem [{\citenamefont {Nandkishore}\ and\ \citenamefont {Huse}(2015)}]{Nandkishore:2014}%
  \BibitemOpen
  \bibfield  {author} {\bibinfo {author} {\bibfnamefont {R.}~\bibnamefont {Nandkishore}}\ and\ \bibinfo {author} {\bibfnamefont {D.~A.}\ \bibnamefont {Huse}},\ }\bibfield  {title} {\bibinfo {title} {{Many body localization and thermalization in quantum statistical mechanics}},\ }\href {https://doi.org/10.1146/annurev-conmatphys-031214-014726} {\bibfield  {journal} {\bibinfo  {journal} {Ann. Rev. Condensed Matter Phys.}\ }\textbf {\bibinfo {volume} {6}},\ \bibinfo {pages} {15} (\bibinfo {year} {2015})},\ \Eprint {https://arxiv.org/abs/1404.0686} {arXiv:1404.0686} \BibitemShut {NoStop}%
\bibitem [{\citenamefont {Altman}\ and\ \citenamefont {Vosk}(2015)}]{Altman:2014}%
  \BibitemOpen
  \bibfield  {author} {\bibinfo {author} {\bibfnamefont {E.}~\bibnamefont {Altman}}\ and\ \bibinfo {author} {\bibfnamefont {R.}~\bibnamefont {Vosk}},\ }\bibfield  {title} {\bibinfo {title} {{Universal dynamics and renormalization in many body localized systems}},\ }\href {https://doi.org/10.1146/annurev-conmatphys-031214-014701} {\bibfield  {journal} {\bibinfo  {journal} {Ann. Rev. Condensed Matter Phys.}\ }\textbf {\bibinfo {volume} {6}},\ \bibinfo {pages} {383} (\bibinfo {year} {2015})},\ \Eprint {https://arxiv.org/abs/1408.2834} {arXiv:1408.2834} \BibitemShut {NoStop}%
\bibitem [{\citenamefont {Abanin}\ \emph {et~al.}(2019)\citenamefont {Abanin}, \citenamefont {Altman}, \citenamefont {Bloch},\ and\ \citenamefont {Serbyn}}]{Abanin:2018}%
  \BibitemOpen
  \bibfield  {author} {\bibinfo {author} {\bibfnamefont {D.~A.}\ \bibnamefont {Abanin}}, \bibinfo {author} {\bibfnamefont {E.}~\bibnamefont {Altman}}, \bibinfo {author} {\bibfnamefont {I.}~\bibnamefont {Bloch}},\ and\ \bibinfo {author} {\bibfnamefont {M.}~\bibnamefont {Serbyn}},\ }\bibfield  {title} {\bibinfo {title} {{Colloquium : Many-body localization, thermalization, and entanglement}},\ }\href {https://doi.org/10.1103/revmodphys.91.021001} {\bibfield  {journal} {\bibinfo  {journal} {Rev. Mod. Phys.}\ }\textbf {\bibinfo {volume} {91}},\ \bibinfo {pages} {021001} (\bibinfo {year} {2019})},\ \Eprint {https://arxiv.org/abs/1804.11065} {arXiv:1804.11065} \BibitemShut {NoStop}%
\bibitem [{\citenamefont {Serbyn}\ \emph {et~al.}(2021)\citenamefont {Serbyn}, \citenamefont {Abanin},\ and\ \citenamefont {Papi{\'c}}}]{Serbyn:2020}%
  \BibitemOpen
  \bibfield  {author} {\bibinfo {author} {\bibfnamefont {M.}~\bibnamefont {Serbyn}}, \bibinfo {author} {\bibfnamefont {D.~A.}\ \bibnamefont {Abanin}},\ and\ \bibinfo {author} {\bibfnamefont {Z.}~\bibnamefont {Papi{\'c}}},\ }\bibfield  {title} {\bibinfo {title} {{Quantum many-body scars and weak breaking of ergodicity}},\ }\href {https://doi.org/10.1038/s41567-021-01230-2} {\bibfield  {journal} {\bibinfo  {journal} {Nature Phys.}\ }\textbf {\bibinfo {volume} {17}},\ \bibinfo {pages} {675} (\bibinfo {year} {2021})},\ \Eprint {https://arxiv.org/abs/2011.09486} {arXiv:2011.09486} \BibitemShut {NoStop}%
\bibitem [{\citenamefont {Chandran}\ \emph {et~al.}(2023)\citenamefont {Chandran}, \citenamefont {Iadecola}, \citenamefont {Khemani},\ and\ \citenamefont {Moessner}}]{Chandran:2022}%
  \BibitemOpen
  \bibfield  {author} {\bibinfo {author} {\bibfnamefont {A.}~\bibnamefont {Chandran}}, \bibinfo {author} {\bibfnamefont {T.}~\bibnamefont {Iadecola}}, \bibinfo {author} {\bibfnamefont {V.}~\bibnamefont {Khemani}},\ and\ \bibinfo {author} {\bibfnamefont {R.}~\bibnamefont {Moessner}},\ }\bibfield  {title} {\bibinfo {title} {{Quantum many-body scars: A quasiparticle perspective}},\ }\href {https://doi.org/10.1146/annurev-conmatphys-031620-101617} {\bibfield  {journal} {\bibinfo  {journal} {Ann. Rev. Condensed Matter Phys.}\ }\textbf {\bibinfo {volume} {14}},\ \bibinfo {pages} {443} (\bibinfo {year} {2023})},\ \Eprint {https://arxiv.org/abs/2206.11528} {arXiv:2206.11528} \BibitemShut {NoStop}%
\bibitem [{\citenamefont {Moudgalya}\ \emph {et~al.}(2022)\citenamefont {Moudgalya}, \citenamefont {Bernevig},\ and\ \citenamefont {Regnault}}]{Moudgalya:2021}%
  \BibitemOpen
  \bibfield  {author} {\bibinfo {author} {\bibfnamefont {S.}~\bibnamefont {Moudgalya}}, \bibinfo {author} {\bibfnamefont {B.~A.}\ \bibnamefont {Bernevig}},\ and\ \bibinfo {author} {\bibfnamefont {N.}~\bibnamefont {Regnault}},\ }\bibfield  {title} {\bibinfo {title} {{Quantum many-body scars and Hilbert space fragmentation: a review of exact results}},\ }\href {https://doi.org/10.1088/1361-6633/ac73a0} {\bibfield  {journal} {\bibinfo  {journal} {Rept. Prog. Phys.}\ }\textbf {\bibinfo {volume} {85}},\ \bibinfo {pages} {086501} (\bibinfo {year} {2022})},\ \Eprint {https://arxiv.org/abs/2109.00548} {arXiv:2109.00548} \BibitemShut {NoStop}%
\bibitem [{\citenamefont {Sala}\ \emph {et~al.}(2020)\citenamefont {Sala}, \citenamefont {Rakovszky}, \citenamefont {Verresen}, \citenamefont {Knap},\ and\ \citenamefont {Pollmann}}]{Sala:2020}%
  \BibitemOpen
  \bibfield  {author} {\bibinfo {author} {\bibfnamefont {P.}~\bibnamefont {Sala}}, \bibinfo {author} {\bibfnamefont {T.}~\bibnamefont {Rakovszky}}, \bibinfo {author} {\bibfnamefont {R.}~\bibnamefont {Verresen}}, \bibinfo {author} {\bibfnamefont {M.}~\bibnamefont {Knap}},\ and\ \bibinfo {author} {\bibfnamefont {F.}~\bibnamefont {Pollmann}},\ }\bibfield  {title} {\bibinfo {title} {{Ergodicity-breaking arising from Hilbert space fragmentation in dipole-conserving Hamiltonians}},\ }\href {https://doi.org/10.1103/PhysRevX.10.011047} {\bibfield  {journal} {\bibinfo  {journal} {Phys. Rev. X}\ }\textbf {\bibinfo {volume} {10}},\ \bibinfo {pages} {011047} (\bibinfo {year} {2020})},\ \Eprint {https://arxiv.org/abs/1904.04266} {arXiv:1904.04266} \BibitemShut {NoStop}%
\bibitem [{\citenamefont {Khemani}\ \emph {et~al.}(2020)\citenamefont {Khemani}, \citenamefont {Hermele},\ and\ \citenamefont {Nandkishore}}]{Khemani:2019}%
  \BibitemOpen
  \bibfield  {author} {\bibinfo {author} {\bibfnamefont {V.}~\bibnamefont {Khemani}}, \bibinfo {author} {\bibfnamefont {M.}~\bibnamefont {Hermele}},\ and\ \bibinfo {author} {\bibfnamefont {R.}~\bibnamefont {Nandkishore}},\ }\bibfield  {title} {\bibinfo {title} {{Localization from Hilbert space shattering: From theory to physical realizations}},\ }\href {https://doi.org/10.1103/PhysRevB.101.174204} {\bibfield  {journal} {\bibinfo  {journal} {Phys. Rev. B}\ }\textbf {\bibinfo {volume} {101}},\ \bibinfo {pages} {174204} (\bibinfo {year} {2020})},\ \Eprint {https://arxiv.org/abs/1904.04815} {arXiv:1904.04815} \BibitemShut {NoStop}%
\bibitem [{\citenamefont {Moudgalya}\ and\ \citenamefont {Motrunich}(2022)}]{Motrunich:2021}%
  \BibitemOpen
  \bibfield  {author} {\bibinfo {author} {\bibfnamefont {S.}~\bibnamefont {Moudgalya}}\ and\ \bibinfo {author} {\bibfnamefont {O.~I.}\ \bibnamefont {Motrunich}},\ }\bibfield  {title} {\bibinfo {title} {{Hilbert space fragmentation and commutant algebras}},\ }\href {https://doi.org/10.1103/PhysRevX.12.011050} {\bibfield  {journal} {\bibinfo  {journal} {Phys. Rev. X}\ }\textbf {\bibinfo {volume} {12}},\ \bibinfo {pages} {011050} (\bibinfo {year} {2022})},\ \Eprint {https://arxiv.org/abs/2108.10324} {arXiv:2108.10324} \BibitemShut {NoStop}%
\bibitem [{\citenamefont {Maldacena}\ \emph {et~al.}(2016)\citenamefont {Maldacena}, \citenamefont {Shenker},\ and\ \citenamefont {Stanford}}]{MSS}%
  \BibitemOpen
  \bibfield  {author} {\bibinfo {author} {\bibfnamefont {J.}~\bibnamefont {Maldacena}}, \bibinfo {author} {\bibfnamefont {S.~H.}\ \bibnamefont {Shenker}},\ and\ \bibinfo {author} {\bibfnamefont {D.}~\bibnamefont {Stanford}},\ }\bibfield  {title} {\bibinfo {title} {{A bound on chaos}},\ }\href {https://doi.org/10.1007/JHEP08(2016)106} {\bibfield  {journal} {\bibinfo  {journal} {J. High Energy Phys.}\ }\textbf {\bibinfo {volume} {2016}}\bibfield  {number} {\bibinfo  {number} { (08)},\ \bibinfo {pages} {106}},\ }\Eprint {https://arxiv.org/abs/1503.01409} {arXiv:1503.01409} \BibitemShut {NoStop}%
\bibitem [{\citenamefont {Swingle}(2018)}]{Swingle:2018}%
  \BibitemOpen
  \bibfield  {author} {\bibinfo {author} {\bibfnamefont {B.}~\bibnamefont {Swingle}},\ }\bibfield  {title} {\bibinfo {title} {{Unscrambling the physics of out-of-time-order correlators}},\ }\href {https://doi.org/10.1038/s41567-018-0295-5} {\bibfield  {journal} {\bibinfo  {journal} {Nature Phys.}\ }\textbf {\bibinfo {volume} {14}},\ \bibinfo {pages} {988} (\bibinfo {year} {2018})}\BibitemShut {NoStop}%
\bibitem [{\citenamefont {Xu}\ and\ \citenamefont {Swingle}(2024)}]{Xu:2024}%
  \BibitemOpen
  \bibfield  {author} {\bibinfo {author} {\bibfnamefont {S.}~\bibnamefont {Xu}}\ and\ \bibinfo {author} {\bibfnamefont {B.}~\bibnamefont {Swingle}},\ }\bibfield  {title} {\bibinfo {title} {{Scrambling dynamics and out-of-time-ordered correlators in quantum many-body systems}},\ }\href {https://doi.org/10.1103/PRXQuantum.5.010201} {\bibfield  {journal} {\bibinfo  {journal} {PRX Quantum}\ }\textbf {\bibinfo {volume} {5}},\ \bibinfo {pages} {010201} (\bibinfo {year} {2024})},\ \Eprint {https://arxiv.org/abs/2202.07060} {arXiv:2202.07060} \BibitemShut {NoStop}%
\bibitem [{\citenamefont {Parker}\ \emph {et~al.}(2019)\citenamefont {Parker}, \citenamefont {Cao}, \citenamefont {Avdoshkin}, \citenamefont {Scaffidi},\ and\ \citenamefont {Altman}}]{Parker:2018}%
  \BibitemOpen
  \bibfield  {author} {\bibinfo {author} {\bibfnamefont {D.~E.}\ \bibnamefont {Parker}}, \bibinfo {author} {\bibfnamefont {X.}~\bibnamefont {Cao}}, \bibinfo {author} {\bibfnamefont {A.}~\bibnamefont {Avdoshkin}}, \bibinfo {author} {\bibfnamefont {T.}~\bibnamefont {Scaffidi}},\ and\ \bibinfo {author} {\bibfnamefont {E.}~\bibnamefont {Altman}},\ }\bibfield  {title} {\bibinfo {title} {{A universal operator growth hypothesis}},\ }\href {https://doi.org/10.1103/PhysRevX.9.041017} {\bibfield  {journal} {\bibinfo  {journal} {Phys. Rev. X}\ }\textbf {\bibinfo {volume} {9}},\ \bibinfo {pages} {041017} (\bibinfo {year} {2019})},\ \Eprint {https://arxiv.org/abs/1812.08657} {arXiv:1812.08657} \BibitemShut {NoStop}%
\bibitem [{\citenamefont {Balasubramanian}\ \emph {et~al.}(2022)\citenamefont {Balasubramanian}, \citenamefont {Caputa}, \citenamefont {Magan},\ and\ \citenamefont {Wu}}]{Balasubramanian:2022-1}%
  \BibitemOpen
  \bibfield  {author} {\bibinfo {author} {\bibfnamefont {V.}~\bibnamefont {Balasubramanian}}, \bibinfo {author} {\bibfnamefont {P.}~\bibnamefont {Caputa}}, \bibinfo {author} {\bibfnamefont {J.~M.}\ \bibnamefont {Magan}},\ and\ \bibinfo {author} {\bibfnamefont {Q.}~\bibnamefont {Wu}},\ }\bibfield  {title} {\bibinfo {title} {{Quantum chaos and the complexity of spread of states}},\ }\href {https://doi.org/10.1103/PhysRevD.106.046007} {\bibfield  {journal} {\bibinfo  {journal} {Phys. Rev. D}\ }\textbf {\bibinfo {volume} {106}},\ \bibinfo {pages} {046007} (\bibinfo {year} {2022})},\ \Eprint {https://arxiv.org/abs/2202.06957} {arXiv:2202.06957} \BibitemShut {NoStop}%
\bibitem [{\citenamefont {Nandy}\ \emph {et~al.}()\citenamefont {Nandy}, \citenamefont {Matsoukas-Roubeas}, \citenamefont {Mart{\'\i}nez-Azcona}, \citenamefont {Dymarsky},\ and\ \citenamefont {del Campo}}]{Dymarsky:2024}%
  \BibitemOpen
  \bibfield  {author} {\bibinfo {author} {\bibfnamefont {P.}~\bibnamefont {Nandy}}, \bibinfo {author} {\bibfnamefont {A.~S.}\ \bibnamefont {Matsoukas-Roubeas}}, \bibinfo {author} {\bibfnamefont {P.}~\bibnamefont {Mart{\'\i}nez-Azcona}}, \bibinfo {author} {\bibfnamefont {A.}~\bibnamefont {Dymarsky}},\ and\ \bibinfo {author} {\bibfnamefont {A.}~\bibnamefont {del Campo}},\ }\bibfield  {title} {\bibinfo {title} {{Quantum dynamics in Krylov space: methods and applications}},\ }\Eprint {https://arxiv.org/abs/2405.09628} {arXiv:2405.09628} \BibitemShut {NoStop}%
\bibitem [{\citenamefont {Viswanath}\ and\ \citenamefont {M{\"u}ller}(1994)}]{Viswanath:book}%
  \BibitemOpen
  \bibfield  {author} {\bibinfo {author} {\bibfnamefont {V.~S.}\ \bibnamefont {Viswanath}}\ and\ \bibinfo {author} {\bibfnamefont {G.}~\bibnamefont {M{\"u}ller}},\ }\href {https://doi.org/10.1007/978-3-540-48651-0} {\emph {\bibinfo {title} {{The Recursion Method: Applications to Many-Body Dynamics}}}},\ Lecture Notes in Physics Monographs\ (\bibinfo  {publisher} {Springer-Verlag},\ \bibinfo {address} {Berlin},\ \bibinfo {year} {1994})\BibitemShut {NoStop}%
\bibitem [{\citenamefont {Motta}\ \emph {et~al.}(2019)\citenamefont {Motta} \emph {et~al.}}]{Motta:2019}%
  \BibitemOpen
  \bibfield  {author} {\bibinfo {author} {\bibfnamefont {M.}~\bibnamefont {Motta}} \emph {et~al.},\ }\bibfield  {title} {\bibinfo {title} {{Determining eigenstates and thermal states on a quantum computer using quantum imaginary time evolution}},\ }\href {https://doi.org/10.1038/s41567-019-0704-4} {\bibfield  {journal} {\bibinfo  {journal} {Nature Phys.}\ }\textbf {\bibinfo {volume} {16}},\ \bibinfo {pages} {205} (\bibinfo {year} {2019})},\ \Eprint {https://arxiv.org/abs/1901.07653} {arXiv:1901.07653} \BibitemShut {NoStop}%
\bibitem [{\citenamefont {Stair}\ \emph {et~al.}(2020)\citenamefont {Stair}, \citenamefont {Huang},\ and\ \citenamefont {Evangelista}}]{Stair:2019}%
  \BibitemOpen
  \bibfield  {author} {\bibinfo {author} {\bibfnamefont {N.~H.}\ \bibnamefont {Stair}}, \bibinfo {author} {\bibfnamefont {R.}~\bibnamefont {Huang}},\ and\ \bibinfo {author} {\bibfnamefont {F.~A.}\ \bibnamefont {Evangelista}},\ }\bibfield  {title} {\bibinfo {title} {{A multireference quantum Krylov algorithm for strongly correlated electrons}},\ }\href {https://doi.org/10.1021/acs.jctc.9b01125} {\bibfield  {journal} {\bibinfo  {journal} {J. Chem. Theory Comput.}\ }\textbf {\bibinfo {volume} {16}},\ \bibinfo {pages} {2236} (\bibinfo {year} {2020})},\ \Eprint {https://arxiv.org/abs/1911.05163} {arXiv:1911.05163} \BibitemShut {NoStop}%
\bibitem [{\citenamefont {Cortes}\ and\ \citenamefont {Gray}(2022)}]{Cortes:2022}%
  \BibitemOpen
  \bibfield  {author} {\bibinfo {author} {\bibfnamefont {C.~L.}\ \bibnamefont {Cortes}}\ and\ \bibinfo {author} {\bibfnamefont {S.~K.}\ \bibnamefont {Gray}},\ }\bibfield  {title} {\bibinfo {title} {{Quantum Krylov subspace algorithms for ground- and excited-state energy estimation}},\ }\href {https://doi.org/10.1103/PhysRevA.105.022417} {\bibfield  {journal} {\bibinfo  {journal} {Phys. Rev. A 105}\ }\textbf {\bibinfo {volume} {105}},\ \bibinfo {pages} {022417} (\bibinfo {year} {2022})},\ \Eprint {https://arxiv.org/abs/2109.06868} {arXiv:2109.06868} \BibitemShut {NoStop}%
\bibitem [{\citenamefont {Larocca}\ and\ \citenamefont {Wisniacki}(2021)}]{Larocca:2021}%
  \BibitemOpen
  \bibfield  {author} {\bibinfo {author} {\bibfnamefont {M.}~\bibnamefont {Larocca}}\ and\ \bibinfo {author} {\bibfnamefont {D.}~\bibnamefont {Wisniacki}},\ }\bibfield  {title} {\bibinfo {title} {{Krylov-subspace approach for the efficient control of quantum many-body dynamics}},\ }\href {https://doi.org/10.1103/PhysRevA.103.023107} {\bibfield  {journal} {\bibinfo  {journal} {Phys. Rev. A 105}\ }\textbf {\bibinfo {volume} {103}},\ \bibinfo {pages} {023107} (\bibinfo {year} {2021})},\ \Eprint {https://arxiv.org/abs/2010.03598} {arXiv:2010.03598} \BibitemShut {NoStop}%
\bibitem [{\citenamefont {Takahashi}\ and\ \citenamefont {del Campo}(2024)}]{Takahashi:2023}%
  \BibitemOpen
  \bibfield  {author} {\bibinfo {author} {\bibfnamefont {K.}~\bibnamefont {Takahashi}}\ and\ \bibinfo {author} {\bibfnamefont {A.}~\bibnamefont {del Campo}},\ }\bibfield  {title} {\bibinfo {title} {{Shortcuts to adiabaticity in Krylov space}},\ }\href {https://doi.org/10.1103/PhysRevX.14.011032} {\bibfield  {journal} {\bibinfo  {journal} {Phys. Rev. X}\ }\textbf {\bibinfo {volume} {14}},\ \bibinfo {pages} {011032} (\bibinfo {year} {2024})},\ \Eprint {https://arxiv.org/abs/2302.05460} {arXiv:2302.05460} \BibitemShut {NoStop}%
\bibitem [{\citenamefont {Yates}\ \emph {et~al.}(2020)\citenamefont {Yates}, \citenamefont {Abanov},\ and\ \citenamefont {Mitra}}]{Yates:2020}%
  \BibitemOpen
  \bibfield  {author} {\bibinfo {author} {\bibfnamefont {D.~J.}\ \bibnamefont {Yates}}, \bibinfo {author} {\bibfnamefont {A.~G.}\ \bibnamefont {Abanov}},\ and\ \bibinfo {author} {\bibfnamefont {A.}~\bibnamefont {Mitra}},\ }\bibfield  {title} {\bibinfo {title} {{Lifetime of almost strong edge-mode operators in one dimensional, interacting, symmetry protected topological phases}},\ }\href {https://doi.org/10.1103/PhysRevLett.124.206803} {\bibfield  {journal} {\bibinfo  {journal} {Phys. Rev. Lett.}\ }\textbf {\bibinfo {volume} {124}},\ \bibinfo {pages} {206803} (\bibinfo {year} {2020})},\ \Eprint {https://arxiv.org/abs/2002.00098} {arXiv:2002.00098} \BibitemShut {NoStop}%
\bibitem [{\citenamefont {Caputa}\ and\ \citenamefont {Liu}(2022)}]{Caputa:2022-1}%
  \BibitemOpen
  \bibfield  {author} {\bibinfo {author} {\bibfnamefont {P.}~\bibnamefont {Caputa}}\ and\ \bibinfo {author} {\bibfnamefont {S.}~\bibnamefont {Liu}},\ }\bibfield  {title} {\bibinfo {title} {{Quantum complexity and topological phases of matter}},\ }\href {https://doi.org/10.1103/PhysRevB.106.195125} {\bibfield  {journal} {\bibinfo  {journal} {Phys. Rev. B}\ }\textbf {\bibinfo {volume} {106}},\ \bibinfo {pages} {195125} (\bibinfo {year} {2022})},\ \Eprint {https://arxiv.org/abs/2205.05688} {arXiv:2205.05688} \BibitemShut {NoStop}%
\bibitem [{\citenamefont {Caputa}\ \emph {et~al.}(2023)\citenamefont {Caputa}, \citenamefont {Gupta}, \citenamefont {Haque}, \citenamefont {Liu}, \citenamefont {Murugan},\ and\ \citenamefont {Van~Zyl}}]{Caputa:2022-2}%
  \BibitemOpen
  \bibfield  {author} {\bibinfo {author} {\bibfnamefont {P.}~\bibnamefont {Caputa}}, \bibinfo {author} {\bibfnamefont {N.}~\bibnamefont {Gupta}}, \bibinfo {author} {\bibfnamefont {S.~S.}\ \bibnamefont {Haque}}, \bibinfo {author} {\bibfnamefont {S.}~\bibnamefont {Liu}}, \bibinfo {author} {\bibfnamefont {J.}~\bibnamefont {Murugan}},\ and\ \bibinfo {author} {\bibfnamefont {H.~J.~R.}\ \bibnamefont {Van~Zyl}},\ }\bibfield  {title} {\bibinfo {title} {{Spread complexity and topological transitions in the Kitaev chain}},\ }\href {https://doi.org/10.1007/JHEP01(2023)120} {\bibfield  {journal} {\bibinfo  {journal} {J. High Energy Phys.}\ }\textbf {\bibinfo {volume} {2023}}\bibfield  {number} {\bibinfo  {number} { (01)},\ \bibinfo {pages} {120}},\ }\Eprint {https://arxiv.org/abs/2208.06311} {arXiv:2208.06311} \BibitemShut {NoStop}%
\bibitem [{\citenamefont {Bhattacharjee}\ \emph {et~al.}(2022{\natexlab{a}})\citenamefont {Bhattacharjee}, \citenamefont {Sur},\ and\ \citenamefont {Nandy}}]{Bhattacharjee:2022-MBL}%
  \BibitemOpen
  \bibfield  {author} {\bibinfo {author} {\bibfnamefont {B.}~\bibnamefont {Bhattacharjee}}, \bibinfo {author} {\bibfnamefont {S.}~\bibnamefont {Sur}},\ and\ \bibinfo {author} {\bibfnamefont {P.}~\bibnamefont {Nandy}},\ }\bibfield  {title} {\bibinfo {title} {{Probing quantum scars and weak ergodicity breaking through quantum complexity}},\ }\href {https://doi.org/10.1103/PhysRevB.106.205150} {\bibfield  {journal} {\bibinfo  {journal} {Phys. Rev. B}\ }\textbf {\bibinfo {volume} {106}},\ \bibinfo {pages} {205150} (\bibinfo {year} {2022}{\natexlab{a}})},\ \Eprint {https://arxiv.org/abs/2208.05503} {arXiv:2208.05503} \BibitemShut {NoStop}%
\bibitem [{\citenamefont {Bhattacharjee}\ \emph {et~al.}(2022{\natexlab{b}})\citenamefont {Bhattacharjee}, \citenamefont {Cao}, \citenamefont {Nandy},\ and\ \citenamefont {Pathak}}]{Bhattacharjee:2022}%
  \BibitemOpen
  \bibfield  {author} {\bibinfo {author} {\bibfnamefont {B.}~\bibnamefont {Bhattacharjee}}, \bibinfo {author} {\bibfnamefont {X.}~\bibnamefont {Cao}}, \bibinfo {author} {\bibfnamefont {P.}~\bibnamefont {Nandy}},\ and\ \bibinfo {author} {\bibfnamefont {T.}~\bibnamefont {Pathak}},\ }\bibfield  {title} {\bibinfo {title} {{Krylov complexity in saddle-dominated scrambling}},\ }\href {https://doi.org/10.1007/JHEP05(2022)174} {\bibfield  {journal} {\bibinfo  {journal} {J. High Energy Phys.}\ }\textbf {\bibinfo {volume} {2022}}\bibfield  {number} {\bibinfo  {number} { (05)},\ \bibinfo {pages} {174}},\ }\Eprint {https://arxiv.org/abs/2203.03534} {arXiv:2203.03534} \BibitemShut {NoStop}%
\bibitem [{\citenamefont {Hashimoto}\ \emph {et~al.}(2023)\citenamefont {Hashimoto}, \citenamefont {Murata}, \citenamefont {Tanahashi},\ and\ \citenamefont {Watanabe}}]{Hashimoto:2023}%
  \BibitemOpen
  \bibfield  {author} {\bibinfo {author} {\bibfnamefont {K.}~\bibnamefont {Hashimoto}}, \bibinfo {author} {\bibfnamefont {K.}~\bibnamefont {Murata}}, \bibinfo {author} {\bibfnamefont {N.}~\bibnamefont {Tanahashi}},\ and\ \bibinfo {author} {\bibfnamefont {R.}~\bibnamefont {Watanabe}},\ }\bibfield  {title} {\bibinfo {title} {{Krylov complexity and chaos in quantum mechanics}},\ }\href {https://doi.org/10.1007/JHEP11(2023)040} {\bibfield  {journal} {\bibinfo  {journal} {J. High Energy Phys.}\ }\textbf {\bibinfo {volume} {2023}}\bibfield  {number} {\bibinfo  {number} { (11)},\ \bibinfo {pages} {040}},\ }\Eprint {https://arxiv.org/abs/2305.16669} {arXiv:2305.16669} \BibitemShut {NoStop}%
\bibitem [{\citenamefont {Dymarsky}\ and\ \citenamefont {Smolkin}(2021)}]{Dymarsky:2021}%
  \BibitemOpen
  \bibfield  {author} {\bibinfo {author} {\bibfnamefont {A.}~\bibnamefont {Dymarsky}}\ and\ \bibinfo {author} {\bibfnamefont {M.}~\bibnamefont {Smolkin}},\ }\bibfield  {title} {\bibinfo {title} {{Krylov complexity in conformal field theory}},\ }\href {https://doi.org/10.1103/PhysRevD.104.L081702} {\bibfield  {journal} {\bibinfo  {journal} {Phys. Rev. D}\ }\textbf {\bibinfo {volume} {104}},\ \bibinfo {pages} {L081702} (\bibinfo {year} {2021})},\ \Eprint {https://arxiv.org/abs/2104.09514} {arXiv:2104.09514} \BibitemShut {NoStop}%
\bibitem [{\citenamefont {Avdoshkin}\ \emph {et~al.}(2024)\citenamefont {Avdoshkin}, \citenamefont {Dymarsky},\ and\ \citenamefont {Smolkin}}]{Dymarsky:2022}%
  \BibitemOpen
  \bibfield  {author} {\bibinfo {author} {\bibfnamefont {A.}~\bibnamefont {Avdoshkin}}, \bibinfo {author} {\bibfnamefont {A.}~\bibnamefont {Dymarsky}},\ and\ \bibinfo {author} {\bibfnamefont {M.}~\bibnamefont {Smolkin}},\ }\bibfield  {title} {\bibinfo {title} {{Krylov complexity in quantum field theory, and beyond}},\ }\href {https://doi.org/10.1007/JHEP06(2024)066} {\bibfield  {journal} {\bibinfo  {journal} {J. High Energy Phys.}\ }\textbf {\bibinfo {volume} {2024}}\bibfield  {number} {\bibinfo  {number} { (06)},\ \bibinfo {pages} {066}},\ }\Eprint {https://arxiv.org/abs/2212.14429} {arXiv:2212.14429} \BibitemShut {NoStop}%
\bibitem [{\citenamefont {Barb{\'o}n}\ \emph {et~al.}(2019)\citenamefont {Barb{\'o}n}, \citenamefont {Rabinovici}, \citenamefont {Shir},\ and\ \citenamefont {Sinha}}]{Rabinovici:2019}%
  \BibitemOpen
  \bibfield  {author} {\bibinfo {author} {\bibfnamefont {J.~L.~F.}\ \bibnamefont {Barb{\'o}n}}, \bibinfo {author} {\bibfnamefont {E.}~\bibnamefont {Rabinovici}}, \bibinfo {author} {\bibfnamefont {R.}~\bibnamefont {Shir}},\ and\ \bibinfo {author} {\bibfnamefont {R.}~\bibnamefont {Sinha}},\ }\bibfield  {title} {\bibinfo {title} {{On The evolution of operator complexity beyond scrambling}},\ }\href {https://doi.org/10.1007/JHEP10(2019)264} {\bibfield  {journal} {\bibinfo  {journal} {J. High Energy Phys.}\ }\textbf {\bibinfo {volume} {2019}}\bibfield  {number} {\bibinfo  {number} { (10)},\ \bibinfo {pages} {264}},\ }\Eprint {https://arxiv.org/abs/1907.05393} {arXiv:1907.05393} \BibitemShut {NoStop}%
\bibitem [{\citenamefont {Rabinovici}\ \emph {et~al.}(2021)\citenamefont {Rabinovici}, \citenamefont {S{\'a}nchez-Garrido}, \citenamefont {Shir},\ and\ \citenamefont {Sonner}}]{Rabinovici:2020}%
  \BibitemOpen
  \bibfield  {author} {\bibinfo {author} {\bibfnamefont {E.}~\bibnamefont {Rabinovici}}, \bibinfo {author} {\bibfnamefont {A.}~\bibnamefont {S{\'a}nchez-Garrido}}, \bibinfo {author} {\bibfnamefont {R.}~\bibnamefont {Shir}},\ and\ \bibinfo {author} {\bibfnamefont {J.}~\bibnamefont {Sonner}},\ }\bibfield  {title} {\bibinfo {title} {{Operator complexity: a journey to the edge of Krylov space}},\ }\href {https://doi.org/10.1007/JHEP06(2021)062} {\bibfield  {journal} {\bibinfo  {journal} {J. High Energy Phys.}\ }\textbf {\bibinfo {volume} {2021}}\bibfield  {number} {\bibinfo  {number} { (06)},\ \bibinfo {pages} {062}},\ }\Eprint {https://arxiv.org/abs/2009.01862} {arXiv:2009.01862} \BibitemShut {NoStop}%
\bibitem [{\citenamefont {Rabinovici}\ \emph {et~al.}(2022{\natexlab{a}})\citenamefont {Rabinovici}, \citenamefont {S{\'a}nchez-Garrido}, \citenamefont {Shir},\ and\ \citenamefont {Sonner}}]{Rabinovici:2022}%
  \BibitemOpen
  \bibfield  {author} {\bibinfo {author} {\bibfnamefont {E.}~\bibnamefont {Rabinovici}}, \bibinfo {author} {\bibfnamefont {A.}~\bibnamefont {S{\'a}nchez-Garrido}}, \bibinfo {author} {\bibfnamefont {R.}~\bibnamefont {Shir}},\ and\ \bibinfo {author} {\bibfnamefont {J.}~\bibnamefont {Sonner}},\ }\bibfield  {title} {\bibinfo {title} {{Krylov complexity from integrability to chaos}},\ }\href {https://doi.org/10.1007/JHEP07(2022)151} {\bibfield  {journal} {\bibinfo  {journal} {J. High Energy Phys.}\ }\textbf {\bibinfo {volume} {2022}}\bibfield  {number} {\bibinfo  {number} { (07)},\ \bibinfo {pages} {151}},\ }\Eprint {https://arxiv.org/abs/2207.07701} {arXiv:2207.07701} \BibitemShut {NoStop}%
\bibitem [{\citenamefont {Rabinovici}\ \emph {et~al.}(2023)\citenamefont {Rabinovici}, \citenamefont {S{\'a}nchez-Garrido}, \citenamefont {Shir},\ and\ \citenamefont {Sonner}}]{Rabinovici:2023}%
  \BibitemOpen
  \bibfield  {author} {\bibinfo {author} {\bibfnamefont {E.}~\bibnamefont {Rabinovici}}, \bibinfo {author} {\bibfnamefont {A.}~\bibnamefont {S{\'a}nchez-Garrido}}, \bibinfo {author} {\bibfnamefont {R.}~\bibnamefont {Shir}},\ and\ \bibinfo {author} {\bibfnamefont {J.}~\bibnamefont {Sonner}},\ }\bibfield  {title} {\bibinfo {title} {{A bulk manifestation of Krylov complexity}},\ }\href {https://doi.org/10.1007/JHEP08(2023)213} {\bibfield  {journal} {\bibinfo  {journal} {J. High Energy Phys.}\ }\textbf {\bibinfo {volume} {2023}}\bibfield  {number} {\bibinfo  {number} { (08)},\ \bibinfo {pages} {213}},\ }\Eprint {https://arxiv.org/abs/2305.04355} {arXiv:2305.04355} \BibitemShut {NoStop}%
\bibitem [{\citenamefont {Jian}\ \emph {et~al.}(2021)\citenamefont {Jian}, \citenamefont {Swingle},\ and\ \citenamefont {Xian}}]{Jian:2020}%
  \BibitemOpen
  \bibfield  {author} {\bibinfo {author} {\bibfnamefont {S.-K.}\ \bibnamefont {Jian}}, \bibinfo {author} {\bibfnamefont {B.}~\bibnamefont {Swingle}},\ and\ \bibinfo {author} {\bibfnamefont {Z.-Y.}\ \bibnamefont {Xian}},\ }\bibfield  {title} {\bibinfo {title} {{Complexity growth of operators in the SYK model and in JT gravity}},\ }\href {https://doi.org/10.1007/JHEP03(2021)014} {\bibfield  {journal} {\bibinfo  {journal} {J. High Energy Phys.}\ }\textbf {\bibinfo {volume} {2021}}\bibfield  {number} {\bibinfo  {number} { (03)},\ \bibinfo {pages} {014}},\ }\Eprint {https://arxiv.org/abs/2008.12274} {arXiv:2008.12274} \BibitemShut {NoStop}%
\bibitem [{\citenamefont {Iizuka}\ and\ \citenamefont {Nishida}(2023)}]{Iizuka:2023}%
  \BibitemOpen
  \bibfield  {author} {\bibinfo {author} {\bibfnamefont {N.}~\bibnamefont {Iizuka}}\ and\ \bibinfo {author} {\bibfnamefont {M.}~\bibnamefont {Nishida}},\ }\bibfield  {title} {\bibinfo {title} {{Krylov complexity in the IP matrix model}},\ }\href {https://doi.org/10.1007/JHEP11(2023)065} {\bibfield  {journal} {\bibinfo  {journal} {J. High Energy Phys.}\ }\textbf {\bibinfo {volume} {2023}}\bibfield  {number} {\bibinfo  {number} { (11)},\ \bibinfo {pages} {065}},\ }\Eprint {https://arxiv.org/abs/2306.04805} {arXiv:2306.04805} \BibitemShut {NoStop}%
\bibitem [{\citenamefont {Avdoshkin}\ and\ \citenamefont {Dymarsky}(2020)}]{Avdoshkin:2019}%
  \BibitemOpen
  \bibfield  {author} {\bibinfo {author} {\bibfnamefont {A.}~\bibnamefont {Avdoshkin}}\ and\ \bibinfo {author} {\bibfnamefont {A.}~\bibnamefont {Dymarsky}},\ }\bibfield  {title} {\bibinfo {title} {{Euclidean operator growth and quantum chaos}},\ }\href {https://doi.org/10.1103/PhysRevResearch.2.043234} {\bibfield  {journal} {\bibinfo  {journal} {Phys. Rev. Res.}\ }\textbf {\bibinfo {volume} {2}},\ \bibinfo {pages} {043234} (\bibinfo {year} {2020})},\ \Eprint {https://arxiv.org/abs/1911.09672} {arXiv:1911.09672} \BibitemShut {NoStop}%
\bibitem [{\citenamefont {Caputa}\ \emph {et~al.}(2022)\citenamefont {Caputa}, \citenamefont {Magan},\ and\ \citenamefont {Patramanis}}]{Caputa:2021}%
  \BibitemOpen
  \bibfield  {author} {\bibinfo {author} {\bibfnamefont {P.}~\bibnamefont {Caputa}}, \bibinfo {author} {\bibfnamefont {J.~M.}\ \bibnamefont {Magan}},\ and\ \bibinfo {author} {\bibfnamefont {D.}~\bibnamefont {Patramanis}},\ }\bibfield  {title} {\bibinfo {title} {{Geometry of Krylov complexity}},\ }\href {https://doi.org/10.1103/PhysRevResearch.4.013041} {\bibfield  {journal} {\bibinfo  {journal} {Phys. Rev. Res.}\ }\textbf {\bibinfo {volume} {4}},\ \bibinfo {pages} {013041} (\bibinfo {year} {2022})},\ \Eprint {https://arxiv.org/abs/2109.03824} {arXiv:2109.03824} \BibitemShut {NoStop}%
\bibitem [{\citenamefont {Gorsky}\ \emph {et~al.}(2024)\citenamefont {Gorsky}, \citenamefont {Nechaev},\ and\ \citenamefont {Valov}}]{Gorsky:2024}%
  \BibitemOpen
  \bibfield  {author} {\bibinfo {author} {\bibfnamefont {A.}~\bibnamefont {Gorsky}}, \bibinfo {author} {\bibfnamefont {S.}~\bibnamefont {Nechaev}},\ and\ \bibinfo {author} {\bibfnamefont {A.}~\bibnamefont {Valov}},\ }\bibfield  {title} {\bibinfo {title} {{KPZ scaling from the Krylov space}},\ }\href {https://doi.org/10.1007/JHEP09(2024)021} {\bibfield  {journal} {\bibinfo  {journal} {J. High Energy Phys.}\ }\textbf {\bibinfo {volume} {2024}}\bibfield  {number} {\bibinfo  {number} { (09)},\ \bibinfo {pages} {021}},\ }\Eprint {https://arxiv.org/abs/2406.02782} {arXiv:2406.02782} \BibitemShut {NoStop}%
\bibitem [{\citenamefont {Das}\ \emph {et~al.}()\citenamefont {Das}, \citenamefont {Demulder}, \citenamefont {Erdmenger},\ and\ \citenamefont {Northe}}]{Das:2024}%
  \BibitemOpen
  \bibfield  {author} {\bibinfo {author} {\bibfnamefont {R.~N.}\ \bibnamefont {Das}}, \bibinfo {author} {\bibfnamefont {S.}~\bibnamefont {Demulder}}, \bibinfo {author} {\bibfnamefont {J.}~\bibnamefont {Erdmenger}},\ and\ \bibinfo {author} {\bibfnamefont {C.}~\bibnamefont {Northe}},\ }\bibfield  {title} {\bibinfo {title} {{Spread complexity for the planar limit of holography}},\ }\Eprint {https://arxiv.org/abs/2412.09673} {arXiv:2412.09673} \BibitemShut {NoStop}%
\bibitem [{\citenamefont {Szeg\H{o}}(1939)}]{Szego}%
  \BibitemOpen
  \bibfield  {author} {\bibinfo {author} {\bibfnamefont {G.}~\bibnamefont {Szeg\H{o}}},\ }\href@noop {} {\emph {\bibinfo {title} {Orthogonal Polynomials}}},\ Colloquium Publications, Vol. 23\ (\bibinfo  {publisher} {American Mathematical Society},\ \bibinfo {address} {Providence, RI},\ \bibinfo {year} {1939})\BibitemShut {NoStop}%
\bibitem [{\citenamefont {Simon}(2005)}]{Simon}%
  \BibitemOpen
  \bibfield  {author} {\bibinfo {author} {\bibfnamefont {B.}~\bibnamefont {Simon}},\ }\href@noop {} {\emph {\bibinfo {title} {{Orthogonal Polynomials on the Unit Circle}}}},\ Colloquium Publications, Vol. 54\ (\bibinfo  {publisher} {American Mathematical Society},\ \bibinfo {address} {Providence, RI},\ \bibinfo {year} {2005})\BibitemShut {NoStop}%
\bibitem [{\citenamefont {Ismail}(2005)}]{Ismail:2005}%
  \BibitemOpen
  \bibfield  {author} {\bibinfo {author} {\bibfnamefont {M.~E.~H.}\ \bibnamefont {Ismail}},\ }\href {https://doi.org/10.1017/CBO9781107325982} {\emph {\bibinfo {title} {Classical and Quantum Orthogonal Polynomials in One Variable}}},\ Encyclopedia of Mathematics and its Applications\ (\bibinfo  {publisher} {Cambridge University Press},\ \bibinfo {address} {Cambridge},\ \bibinfo {year} {2005})\BibitemShut {NoStop}%
\bibitem [{\citenamefont {Ismail}\ and\ \citenamefont {Van~Assche}(2020)}]{Ismail:2020}%
  \BibitemOpen
  \bibfield  {author} {\bibinfo {author} {\bibfnamefont {M.~E.~H.}\ \bibnamefont {Ismail}}\ and\ \bibinfo {author} {\bibfnamefont {W.}~\bibnamefont {Van~Assche}},\ }\href {https://doi.org/10.1017/9780511979156} {\emph {\bibinfo {title} {Univariate Orthogonal Polynomials}}},\ Encyclopedia of Special Functions: The Askey-Bateman Project, Vol 1.\ (\bibinfo  {publisher} {Cambridge University Press},\ \bibinfo {address} {Cambridge},\ \bibinfo {year} {2020})\BibitemShut {NoStop}%
\bibitem [{\citenamefont {M\"uck}\ and\ \citenamefont {Yang}(2022)}]{Muck:2022}%
  \BibitemOpen
  \bibfield  {author} {\bibinfo {author} {\bibfnamefont {W.}~\bibnamefont {M\"uck}}\ and\ \bibinfo {author} {\bibfnamefont {Y.}~\bibnamefont {Yang}},\ }\bibfield  {title} {\bibinfo {title} {{Krylov complexity and orthogonal polynomials}},\ }\href {https://doi.org/10.1016/j.nuclphysb.2022.115948} {\bibfield  {journal} {\bibinfo  {journal} {Nucl. Phys. B}\ }\textbf {\bibinfo {volume} {984}},\ \bibinfo {pages} {115948} (\bibinfo {year} {2022})},\ \Eprint {https://arxiv.org/abs/2205.12815} {arXiv:2205.12815} \BibitemShut {NoStop}%
\bibitem [{\citenamefont {Yates}\ \emph {et~al.}(2022)\citenamefont {Yates}, \citenamefont {Abanov},\ and\ \citenamefont {Mitra}}]{Yates:2021b}%
  \BibitemOpen
  \bibfield  {author} {\bibinfo {author} {\bibfnamefont {D.~J.}\ \bibnamefont {Yates}}, \bibinfo {author} {\bibfnamefont {A.~G.}\ \bibnamefont {Abanov}},\ and\ \bibinfo {author} {\bibfnamefont {A.}~\bibnamefont {Mitra}},\ }\bibfield  {title} {\bibinfo {title} {{Long-lived period-doubled edge modes of interacting and disorder-free Floquet spin chains}},\ }\href {https://doi.org/10.1038/s42005-022-00818-1} {\bibfield  {journal} {\bibinfo  {journal} {Commun. Phys.}\ }\textbf {\bibinfo {volume} {5}},\ \bibinfo {pages} {43} (\bibinfo {year} {2022})},\ \Eprint {https://arxiv.org/abs/2105.13766} {arXiv:2105.13766} \BibitemShut {NoStop}%
\bibitem [{\citenamefont {Yates}\ and\ \citenamefont {Mitra}(2021)}]{Yates:2021a}%
  \BibitemOpen
  \bibfield  {author} {\bibinfo {author} {\bibfnamefont {D.~J.}\ \bibnamefont {Yates}}\ and\ \bibinfo {author} {\bibfnamefont {A.}~\bibnamefont {Mitra}},\ }\bibfield  {title} {\bibinfo {title} {{Strong and almost strong modes of Floquet spin chains in Krylov subspaces}},\ }\href {https://doi.org/10.1103/PhysRevB.104.195121} {\bibfield  {journal} {\bibinfo  {journal} {Phys. Rev. B}\ }\textbf {\bibinfo {volume} {104}},\ \bibinfo {pages} {195121} (\bibinfo {year} {2021})},\ \Eprint {https://arxiv.org/abs/2105.13246} {arXiv:2105.13246} \BibitemShut {NoStop}%
\bibitem [{\citenamefont {Qi}\ \emph {et~al.}()\citenamefont {Qi}, \citenamefont {Wu},\ and\ \citenamefont {Zheng}}]{Qi:2024}%
  \BibitemOpen
  \bibfield  {author} {\bibinfo {author} {\bibfnamefont {H.-Y.}\ \bibnamefont {Qi}}, \bibinfo {author} {\bibfnamefont {Y.}~\bibnamefont {Wu}},\ and\ \bibinfo {author} {\bibfnamefont {W.}~\bibnamefont {Zheng}},\ }\bibfield  {title} {\bibinfo {title} {{Topological origin of Floquet thermalization in periodically driven many-body systems}},\ }\Eprint {https://arxiv.org/abs/2404.18052} {arXiv:2404.18052} \BibitemShut {NoStop}%
\bibitem [{\citenamefont {Nizami}\ and\ \citenamefont {Shrestha}(2023)}]{Nizami:2023}%
  \BibitemOpen
  \bibfield  {author} {\bibinfo {author} {\bibfnamefont {A.~A.}\ \bibnamefont {Nizami}}\ and\ \bibinfo {author} {\bibfnamefont {A.~W.}\ \bibnamefont {Shrestha}},\ }\bibfield  {title} {\bibinfo {title} {{Krylov construction and complexity for driven quantum systems}},\ }\href {https://doi.org/10.1103/PhysRevE.108.054222} {\bibfield  {journal} {\bibinfo  {journal} {Phys. Rev. E}\ }\textbf {\bibinfo {volume} {108}},\ \bibinfo {pages} {054222} (\bibinfo {year} {2023})},\ \Eprint {https://arxiv.org/abs/2305.00256} {arXiv:2305.00256} \BibitemShut {NoStop}%
\bibitem [{\citenamefont {Nizami}\ and\ \citenamefont {Shrestha}(2024)}]{Nizami:2024}%
  \BibitemOpen
  \bibfield  {author} {\bibinfo {author} {\bibfnamefont {A.~A.}\ \bibnamefont {Nizami}}\ and\ \bibinfo {author} {\bibfnamefont {A.~W.}\ \bibnamefont {Shrestha}},\ }\bibfield  {title} {\bibinfo {title} {{Spread complexity and quantum chaos for periodically driven spin chains}},\ }\href {https://doi.org/10.1103/PhysRevE.110.034201} {\bibfield  {journal} {\bibinfo  {journal} {Phys. Rev. E}\ }\textbf {\bibinfo {volume} {110}},\ \bibinfo {pages} {034201} (\bibinfo {year} {2024})},\ \Eprint {https://arxiv.org/abs/2405.16182} {arXiv:2405.16182} \BibitemShut {NoStop}%
\bibitem [{\citenamefont {Suchsland}\ \emph {et~al.}()\citenamefont {Suchsland}, \citenamefont {Moessner},\ and\ \citenamefont {Claeys}}]{Suchsland:2023}%
  \BibitemOpen
  \bibfield  {author} {\bibinfo {author} {\bibfnamefont {P.}~\bibnamefont {Suchsland}}, \bibinfo {author} {\bibfnamefont {R.}~\bibnamefont {Moessner}},\ and\ \bibinfo {author} {\bibfnamefont {P.~W.}\ \bibnamefont {Claeys}},\ }\bibfield  {title} {\bibinfo {title} {{Krylov complexity and Trotter transitions in unitary circuit dynamics}},\ }\Eprint {https://arxiv.org/abs/2308.03851} {arXiv:2308.03851} \BibitemShut {NoStop}%
\bibitem [{\citenamefont {Scialchi}\ \emph {et~al.}()\citenamefont {Scialchi}, \citenamefont {Roncaglia}, \citenamefont {Pineda},\ and\ \citenamefont {Wisniacki}}]{Scialchi:2024}%
  \BibitemOpen
  \bibfield  {author} {\bibinfo {author} {\bibfnamefont {G.~F.}\ \bibnamefont {Scialchi}}, \bibinfo {author} {\bibfnamefont {A.~J.}\ \bibnamefont {Roncaglia}}, \bibinfo {author} {\bibfnamefont {C.}~\bibnamefont {Pineda}},\ and\ \bibinfo {author} {\bibfnamefont {D.~A.}\ \bibnamefont {Wisniacki}},\ }\bibfield  {title} {\bibinfo {title} {{Exploring quantum ergodicity of unitary evolution through the Krylov approach}},\ }\Eprint {https://arxiv.org/abs/2407.06428} {arXiv:2407.06428} \BibitemShut {NoStop}%
\bibitem [{\citenamefont {Yeh}\ and\ \citenamefont {Mitra}(2024)}]{Yeh:2023}%
  \BibitemOpen
  \bibfield  {author} {\bibinfo {author} {\bibfnamefont {H.-C.}\ \bibnamefont {Yeh}}\ and\ \bibinfo {author} {\bibfnamefont {A.}~\bibnamefont {Mitra}},\ }\bibfield  {title} {\bibinfo {title} {{Universal model of Floquet operator Krylov space}},\ }\href {https://doi.org/10.1103/PhysRevB.110.155109} {\bibfield  {journal} {\bibinfo  {journal} {Phys. Rev. B}\ }\textbf {\bibinfo {volume} {110}},\ \bibinfo {pages} {155109} (\bibinfo {year} {2024})},\ \Eprint {https://arxiv.org/abs/2311.15116} {arXiv:2311.15116} \BibitemShut {NoStop}%
\bibitem [{\citenamefont {Yeh}\ and\ \citenamefont {Mitra}()}]{Yeh:2024}%
  \BibitemOpen
  \bibfield  {author} {\bibinfo {author} {\bibfnamefont {H.-C.}\ \bibnamefont {Yeh}}\ and\ \bibinfo {author} {\bibfnamefont {A.}~\bibnamefont {Mitra}},\ }\bibfield  {title} {\bibinfo {title} {{Moment method and continued fraction expansion in Floquet Operator Krylov Space}},\ }\Eprint {https://arxiv.org/abs/2410.15223} {arXiv:2410.15223} \BibitemShut {NoStop}%
\bibitem [{Note1()}]{Note1}%
  \BibitemOpen
  \bibinfo {note} {The approach of~\cite {Yeh:2023,Yeh:2024} relies on the peculiarities of operator evolution and always generates real Verblunsky coefficients; so, it constructs a Krylov basis for any operator evolution, but not for any state evolution.}\BibitemShut {Stop}%
\bibitem [{\citenamefont {Dymarsky}\ and\ \citenamefont {Gorsky}(2020)}]{Gorsky:2019}%
  \BibitemOpen
  \bibfield  {author} {\bibinfo {author} {\bibfnamefont {A.}~\bibnamefont {Dymarsky}}\ and\ \bibinfo {author} {\bibfnamefont {A.}~\bibnamefont {Gorsky}},\ }\bibfield  {title} {\bibinfo {title} {{Quantum chaos as delocalization in Krylov space}},\ }\href {https://doi.org/10.1103/PhysRevB.102.085137} {\bibfield  {journal} {\bibinfo  {journal} {Phys. Rev. B}\ }\textbf {\bibinfo {volume} {102}},\ \bibinfo {pages} {085137} (\bibinfo {year} {2020})},\ \Eprint {https://arxiv.org/abs/1912.12227} {arXiv:1912.12227} \BibitemShut {NoStop}%
\bibitem [{\citenamefont {Trigueros}\ and\ \citenamefont {Lin}(2022)}]{Trigueros:2021}%
  \BibitemOpen
  \bibfield  {author} {\bibinfo {author} {\bibfnamefont {F.~B.}\ \bibnamefont {Trigueros}}\ and\ \bibinfo {author} {\bibfnamefont {C.-J.}\ \bibnamefont {Lin}},\ }\bibfield  {title} {\bibinfo {title} {{Krylov complexity of many-body localization: Operator localization in Krylov basis}},\ }\href {https://doi.org/10.21468/SciPostPhys.13.2.037} {\bibfield  {journal} {\bibinfo  {journal} {SciPost Phys.}\ }\textbf {\bibinfo {volume} {13}},\ \bibinfo {pages} {037} (\bibinfo {year} {2022})},\ \Eprint {https://arxiv.org/abs/2112.04722} {arXiv:2112.04722} \BibitemShut {NoStop}%
\bibitem [{\citenamefont {Rabinovici}\ \emph {et~al.}(2022{\natexlab{b}})\citenamefont {Rabinovici}, \citenamefont {S{\'a}nchez-Garrido}, \citenamefont {Shir},\ and\ \citenamefont {Sonner}}]{Rabinovici:2021}%
  \BibitemOpen
  \bibfield  {author} {\bibinfo {author} {\bibfnamefont {E.}~\bibnamefont {Rabinovici}}, \bibinfo {author} {\bibfnamefont {A.}~\bibnamefont {S{\'a}nchez-Garrido}}, \bibinfo {author} {\bibfnamefont {R.}~\bibnamefont {Shir}},\ and\ \bibinfo {author} {\bibfnamefont {J.}~\bibnamefont {Sonner}},\ }\bibfield  {title} {\bibinfo {title} {{Krylov localization and suppression of complexity}},\ }\href {https://doi.org/10.1007/JHEP03(2022)211} {\bibfield  {journal} {\bibinfo  {journal} {J. High Energy Phys.}\ }\textbf {\bibinfo {volume} {2022}}\bibfield  {number} {\bibinfo  {number} { (03)},\ \bibinfo {pages} {211}},\ }\Eprint {https://arxiv.org/abs/2112.12128} {arXiv:2112.12128} \BibitemShut {NoStop}%
\bibitem [{\citenamefont {Menzler}\ and\ \citenamefont {Jha}(2024)}]{Menzler:2024}%
  \BibitemOpen
  \bibfield  {author} {\bibinfo {author} {\bibfnamefont {H.~G.}\ \bibnamefont {Menzler}}\ and\ \bibinfo {author} {\bibfnamefont {R.}~\bibnamefont {Jha}},\ }\bibfield  {title} {\bibinfo {title} {{Krylov delocalization/localization across ergodicity breaking}},\ }\href {https://doi.org/10.1103/PhysRevB.110.125137} {\bibfield  {journal} {\bibinfo  {journal} {Phys. Rev. B}\ }\textbf {\bibinfo {volume} {110}},\ \bibinfo {pages} {125137} (\bibinfo {year} {2024})},\ \Eprint {https://arxiv.org/abs/2403.14384} {arXiv:2403.14384} \BibitemShut {NoStop}%
\bibitem [{\citenamefont {Alaoui}\ and\ \citenamefont {Laburthe-Tolra}(2024)}]{Alaoui:2023}%
  \BibitemOpen
  \bibfield  {author} {\bibinfo {author} {\bibfnamefont {Y.~A.}\ \bibnamefont {Alaoui}}\ and\ \bibinfo {author} {\bibfnamefont {B.}~\bibnamefont {Laburthe-Tolra}},\ }\bibfield  {title} {\bibinfo {title} {{Method to discriminate between localized and chaotic quantum systems}},\ }\href {https://doi.org/10.1103/PhysRevResearch.6.043045} {\bibfield  {journal} {\bibinfo  {journal} {Phys. Rev. Res.}\ }\textbf {\bibinfo {volume} {6}},\ \bibinfo {pages} {043045} (\bibinfo {year} {2024})},\ \Eprint {https://arxiv.org/abs/2307.10706} {arXiv:2307.10706} \BibitemShut {NoStop}%
\bibitem [{\citenamefont {Alishahiha}\ and\ \citenamefont {Vasli}()}]{Alishahiha:2024}%
  \BibitemOpen
  \bibfield  {author} {\bibinfo {author} {\bibfnamefont {M.}~\bibnamefont {Alishahiha}}\ and\ \bibinfo {author} {\bibfnamefont {M.~J.}\ \bibnamefont {Vasli}},\ }\bibfield  {title} {\bibinfo {title} {{Thermalization in Krylov basis}},\ }\Eprint {https://arxiv.org/abs/2403.06655} {arXiv:2403.06655} \BibitemShut {NoStop}%
\bibitem [{Note2()}]{Note2}%
  \BibitemOpen
  \bibinfo {note} {Note that the set of states $| \protect \tilde {\Phi }_n \rangle $ is normalized but not orthogonal: $\langle \protect \tilde {\Phi }_n | \protect \tilde {\Phi }_n \rangle = 1$ but $\langle \protect \tilde {\Phi }_m | \protect \tilde {\Phi }_n \rangle \protect \neq 0$ for $m \protect \neq n$. However, these states are orthogonal to Krylov basis states generated after step $n$: $\langle \Phi _m | \protect \tilde {\Phi }_n \rangle = 0$ for $m > n$. Besides, $ \langle \protect \tilde {\Phi }_m | U | \protect \tilde {\Phi }_n \rangle = 0$ and $ \langle \protect \tilde {\Phi }_m | U | \Phi _n \rangle = 0$ for $m > n$.}\BibitemShut {Stop}%
\bibitem [{\citenamefont {Cantero}\ \emph {et~al.}(2003)\citenamefont {Cantero}, \citenamefont {Moral},\ and\ \citenamefont {Velazquez}}]{CMV}%
  \BibitemOpen
  \bibfield  {author} {\bibinfo {author} {\bibfnamefont {M.~J.}\ \bibnamefont {Cantero}}, \bibinfo {author} {\bibfnamefont {L.}~\bibnamefont {Moral}},\ and\ \bibinfo {author} {\bibfnamefont {L.}~\bibnamefont {Velazquez}},\ }\bibfield  {title} {\bibinfo {title} {{Five-diagonal matrices and zeros of orthogonal polynomials on the unit circle}},\ }\href {https://doi.org/10.1016/S0024-3795(02)00457-3} {\bibfield  {journal} {\bibinfo  {journal} {Linear Algebra Appl.}\ }\textbf {\bibinfo {volume} {362}},\ \bibinfo {pages} {29} (\bibinfo {year} {2003})},\ \Eprint {https://arxiv.org/abs/math/0204300} {arXiv:math/0204300} \BibitemShut {NoStop}%
\bibitem [{Note3()}]{Note3}%
  \BibitemOpen
  \bibinfo {note} {To restrict the dimension of the subspace simulated on the $L$-qubit circuit to $D < 2^L$, one simply needs to set $|\alpha _n| = 1$ for $n \ge D$, which decouples this subspace from the rest of the Hilbert space.}\BibitemShut {Stop}%
\bibitem [{\citenamefont {Shende}\ \emph {et~al.}(2004)\citenamefont {Shende}, \citenamefont {Markov},\ and\ \citenamefont {Bullock}}]{Shende:2004}%
  \BibitemOpen
  \bibfield  {author} {\bibinfo {author} {\bibfnamefont {V.~V.}\ \bibnamefont {Shende}}, \bibinfo {author} {\bibfnamefont {I.~L.}\ \bibnamefont {Markov}},\ and\ \bibinfo {author} {\bibfnamefont {S.~S.}\ \bibnamefont {Bullock}},\ }\bibfield  {title} {\bibinfo {title} {{Minimal universal two-qubit controlled-NOT-based circuits}},\ }\href {https://doi.org/10.1103/PhysRevA.69.062321} {\bibfield  {journal} {\bibinfo  {journal} {Phys. Rev. A}\ }\textbf {\bibinfo {volume} {69}},\ \bibinfo {pages} {062321} (\bibinfo {year} {2004})},\ \Eprint {https://arxiv.org/abs/quant-ph/0308033} {arXiv:quant-ph/0308033} \BibitemShut {NoStop}%
\bibitem [{\citenamefont {Shende}\ \emph {et~al.}(2006)\citenamefont {Shende}, \citenamefont {Bullock},\ and\ \citenamefont {Markov}}]{Shende:2006}%
  \BibitemOpen
  \bibfield  {author} {\bibinfo {author} {\bibfnamefont {V.~V.}\ \bibnamefont {Shende}}, \bibinfo {author} {\bibfnamefont {S.~S.}\ \bibnamefont {Bullock}},\ and\ \bibinfo {author} {\bibfnamefont {I.~L.}\ \bibnamefont {Markov}},\ }\bibfield  {title} {\bibinfo {title} {{Synthesis of quantum-logic circuits}},\ }\href {https://doi.org/10.1109/tcad.2005.855930} {\bibfield  {journal} {\bibinfo  {journal} {IEEE Trans. Comput. Aided Design Integr. Circuits Syst.}\ }\textbf {\bibinfo {volume} {25}},\ \bibinfo {pages} {1000} (\bibinfo {year} {2006})},\ \Eprint {https://arxiv.org/abs/quant-ph/0406176} {arXiv:quant-ph/0406176} \BibitemShut {NoStop}%
\bibitem [{\citenamefont {Killip}\ and\ \citenamefont {Stoiciu}(2009)}]{Killip:2006}%
  \BibitemOpen
  \bibfield  {author} {\bibinfo {author} {\bibfnamefont {R.}~\bibnamefont {Killip}}\ and\ \bibinfo {author} {\bibfnamefont {M.}~\bibnamefont {Stoiciu}},\ }\bibfield  {title} {\bibinfo {title} {{Eigenvalue statistics for CMV matrices: From Poisson to clock via random matrix ensembles}},\ }\href {https://doi.org/10.1215/00127094-2009-001} {\bibfield  {journal} {\bibinfo  {journal} {Duke Math. J.}\ }\textbf {\bibinfo {volume} {146}},\ \bibinfo {pages} {361} (\bibinfo {year} {2009})},\ \Eprint {https://arxiv.org/abs/math-ph/0608002} {arXiv:math-ph/0608002} \BibitemShut {NoStop}%
\bibitem [{\citenamefont {Killip}\ and\ \citenamefont {Nenciu}(2004)}]{Killip:2004}%
  \BibitemOpen
  \bibfield  {author} {\bibinfo {author} {\bibfnamefont {R.}~\bibnamefont {Killip}}\ and\ \bibinfo {author} {\bibfnamefont {I.}~\bibnamefont {Nenciu}},\ }\bibfield  {title} {\bibinfo {title} {{Matrix models for circular ensembles}},\ }\href {https://doi.org/10.1155/S1073792804141597} {\bibfield  {journal} {\bibinfo  {journal} {Int. Math. Res. Not.}\ }\textbf {\bibinfo {volume} {2004}},\ \bibinfo {pages} {2665} (\bibinfo {year} {2004})},\ \Eprint {https://arxiv.org/abs/math/0410034} {arXiv:math/0410034} \BibitemShut {NoStop}%
\bibitem [{\citenamefont {Dumitriu}\ and\ \citenamefont {Edelman}(2002)}]{Dumitriu:2002}%
  \BibitemOpen
  \bibfield  {author} {\bibinfo {author} {\bibfnamefont {I.}~\bibnamefont {Dumitriu}}\ and\ \bibinfo {author} {\bibfnamefont {A.}~\bibnamefont {Edelman}},\ }\bibfield  {title} {\bibinfo {title} {{Matrix models for beta ensembles}},\ }\href {https://doi.org/10.1063/1.1507823} {\bibfield  {journal} {\bibinfo  {journal} {J. Math. Phys.}\ }\textbf {\bibinfo {volume} {43}},\ \bibinfo {pages} {5830} (\bibinfo {year} {2002})},\ \Eprint {https://arxiv.org/abs/math-ph/0206043} {arXiv:math-ph/0206043} \BibitemShut {NoStop}%
\bibitem [{\citenamefont {Das}\ and\ \citenamefont {Ghosh}(2022)}]{Das:2021}%
  \BibitemOpen
  \bibfield  {author} {\bibinfo {author} {\bibfnamefont {A.~K.}\ \bibnamefont {Das}}\ and\ \bibinfo {author} {\bibfnamefont {A.}~\bibnamefont {Ghosh}},\ }\bibfield  {title} {\bibinfo {title} {{Nonergodic extended states in the \ensuremath{\beta} ensemble}},\ }\href {https://doi.org/10.1103/PhysRevE.105.054121} {\bibfield  {journal} {\bibinfo  {journal} {Phys. Rev. E}\ }\textbf {\bibinfo {volume} {105}},\ \bibinfo {pages} {054121} (\bibinfo {year} {2022})},\ \Eprint {https://arxiv.org/abs/2112.11910} {arXiv:2112.11910} \BibitemShut {NoStop}%
\bibitem [{\citenamefont {Balasubramanian}\ \emph {et~al.}(2023)\citenamefont {Balasubramanian}, \citenamefont {Magan},\ and\ \citenamefont {Wu}}]{Balasubramanian:2022-2}%
  \BibitemOpen
  \bibfield  {author} {\bibinfo {author} {\bibfnamefont {V.}~\bibnamefont {Balasubramanian}}, \bibinfo {author} {\bibfnamefont {J.~M.}\ \bibnamefont {Magan}},\ and\ \bibinfo {author} {\bibfnamefont {Q.}~\bibnamefont {Wu}},\ }\bibfield  {title} {\bibinfo {title} {{Tridiagonalizing random matrices}},\ }\href {https://doi.org/10.1103/PhysRevD.107.126001} {\bibfield  {journal} {\bibinfo  {journal} {Phys. Rev. D}\ }\textbf {\bibinfo {volume} {107}},\ \bibinfo {pages} {126001} (\bibinfo {year} {2023})},\ \Eprint {https://arxiv.org/abs/2208.08452} {arXiv:2208.08452} \BibitemShut {NoStop}%
\bibitem [{Note4()}]{Note4}%
  \BibitemOpen
  \bibinfo {note} {For a finite $d$, Verblunsky coefficients might deviate from these distributions, but the average magnitude of deviations goes to zero as $d \to \infty $}\BibitemShut {NoStop}%
\bibitem [{Note5()}]{Note5}%
  \BibitemOpen
  \bibinfo {note} {Note that the eigenvalue distribution of the Krylov chain is only an indirect probe of chaoticity of the original many-body dynamics. The main loophole is that the dimension of the Krylov subspace might be smaller than the total Hilbert space dimension, so it captures only a fraction of energy levels. However, this is not the case for the examples we consider in this Letter.}\BibitemShut {Stop}%
\bibitem [{Note6()}]{Note6}%
  \BibitemOpen
  \bibinfo {note} {Circular $\beta $-ensemble includes the orthogonal ($\beta = 1$), unitary ($\beta = 2$), and symplectic ($\beta = 4$) ensembles.}\BibitemShut {Stop}%
\bibitem [{\citenamefont {Oganesyan}\ and\ \citenamefont {Huse}(2007)}]{Oganesyan:2007}%
  \BibitemOpen
  \bibfield  {author} {\bibinfo {author} {\bibfnamefont {V.}~\bibnamefont {Oganesyan}}\ and\ \bibinfo {author} {\bibfnamefont {D.~A.}\ \bibnamefont {Huse}},\ }\bibfield  {title} {\bibinfo {title} {{Localization of interacting fermions at high temperature}},\ }\href {https://doi.org/10.1103/PhysRevB.75.155111} {\bibfield  {journal} {\bibinfo  {journal} {Phys. Rev. B}\ }\textbf {\bibinfo {volume} {75}},\ \bibinfo {pages} {155111} (\bibinfo {year} {2007})},\ \Eprint {https://arxiv.org/abs/cond-mat/0610854} {arXiv:cond-mat/0610854} \BibitemShut {NoStop}%
\bibitem [{\citenamefont {Atas}\ \emph {et~al.}(2013)\citenamefont {Atas}, \citenamefont {Bogomolny}, \citenamefont {Giraud},\ and\ \citenamefont {Roux}}]{Atas:2013}%
  \BibitemOpen
  \bibfield  {author} {\bibinfo {author} {\bibfnamefont {Y.~Y.}\ \bibnamefont {Atas}}, \bibinfo {author} {\bibfnamefont {E.}~\bibnamefont {Bogomolny}}, \bibinfo {author} {\bibfnamefont {O.}~\bibnamefont {Giraud}},\ and\ \bibinfo {author} {\bibfnamefont {G.}~\bibnamefont {Roux}},\ }\bibfield  {title} {\bibinfo {title} {{Distribution of the ratio of consecutive level spacings in random matrix ensembles}},\ }\href {https://doi.org/10.1103/PhysRevLett.110.084101} {\bibfield  {journal} {\bibinfo  {journal} {Phys. Rev. Lett.}\ }\textbf {\bibinfo {volume} {110}},\ \bibinfo {pages} {084101} (\bibinfo {year} {2013})},\ \Eprint {https://arxiv.org/abs/1212.5611} {arXiv:1212.5611} \BibitemShut {NoStop}%
\bibitem [{\citenamefont {Haake}(2010)}]{Haake}%
  \BibitemOpen
  \bibfield  {author} {\bibinfo {author} {\bibfnamefont {F.}~\bibnamefont {Haake}},\ }\href {https://doi.org/10.1007/978-3-642-05428-0} {\emph {\bibinfo {title} {{Quantum Signatures of Chaos}}}}\ (\bibinfo  {publisher} {Springer-Verlag},\ \bibinfo {address} {Berlin Heidelberg},\ \bibinfo {year} {2010})\BibitemShut {NoStop}%
\bibitem [{\citenamefont {St{\" o}ckmann}(1999)}]{Stockmann}%
  \BibitemOpen
  \bibfield  {author} {\bibinfo {author} {\bibfnamefont {H.-J.}\ \bibnamefont {St{\" o}ckmann}},\ }\href {https://doi.org/10.1017/CBO9780511524622} {\emph {\bibinfo {title} {{Quantum Chaos: An Introduction}}}}\ (\bibinfo  {publisher} {Cambridge University Press},\ \bibinfo {address} {Cambridge, England},\ \bibinfo {year} {1999})\BibitemShut {NoStop}%
\bibitem [{\citenamefont {Berry}\ and\ \citenamefont {Tabor}(1977)}]{Berry:1977a}%
  \BibitemOpen
  \bibfield  {author} {\bibinfo {author} {\bibfnamefont {M.~V.}\ \bibnamefont {Berry}}\ and\ \bibinfo {author} {\bibfnamefont {M.}~\bibnamefont {Tabor}},\ }\bibfield  {title} {\bibinfo {title} {{Level clustering in the regular spectrum}},\ }\href {https://doi.org/10.1098/rspa.1977.0140} {\bibfield  {journal} {\bibinfo  {journal} {Proc. R. Soc. Lond. A}\ }\textbf {\bibinfo {volume} {356}},\ \bibinfo {pages} {375} (\bibinfo {year} {1977})}\BibitemShut {NoStop}%
\bibitem [{\citenamefont {Bohigas}\ \emph {et~al.}(1984)\citenamefont {Bohigas}, \citenamefont {Giannoni},\ and\ \citenamefont {Schmit}}]{Bohigas:1984}%
  \BibitemOpen
  \bibfield  {author} {\bibinfo {author} {\bibfnamefont {O.}~\bibnamefont {Bohigas}}, \bibinfo {author} {\bibfnamefont {M.~J.}\ \bibnamefont {Giannoni}},\ and\ \bibinfo {author} {\bibfnamefont {C.}~\bibnamefont {Schmit}},\ }\bibfield  {title} {\bibinfo {title} {{Characterization of chaotic quantum spectra and universality of level fluctuation laws}},\ }\href {https://doi.org/10.1103/PhysRevLett.52.1} {\bibfield  {journal} {\bibinfo  {journal} {Phys. Rev. Lett.}\ }\textbf {\bibinfo {volume} {52}},\ \bibinfo {pages} {1} (\bibinfo {year} {1984})}\BibitemShut {NoStop}%
\bibitem [{Note7()}]{Note7}%
  \BibitemOpen
  \bibinfo {note} {In fact, eigenvectors of a non-averaged $U$ are localized around different sites and usually have different localization lengths. However, the qualitative prediction $e^S \sim d^{1 - \epsilon }$ re-emerges after averaging over eigenvectors.}\BibitemShut {Stop}%
\bibitem [{\citenamefont {Teplyaev}(1991)}]{Teplyaev:1991}%
  \BibitemOpen
  \bibfield  {author} {\bibinfo {author} {\bibfnamefont {A.~V.}\ \bibnamefont {Teplyaev}},\ }\bibfield  {title} {\bibinfo {title} {{The pure point spectrum of random orthogonal polynomials on the circle}},\ }\href {https://www.mathnet.ru/eng/dan5858} {\bibfield  {journal} {\bibinfo  {journal} {Dokl. Akad. Nauk}\ }\textbf {\bibinfo {volume} {320}},\ \bibinfo {pages} {49} (\bibinfo {year} {1991})},\ \bibinfo {note} {[Dokl. Math. \textbf{44}, 407 (1992)]}\BibitemShut {NoStop}%
\bibitem [{\citenamefont {Cedzich}\ and\ \citenamefont {Werner}(2021)}]{Cedzich:2019}%
  \BibitemOpen
  \bibfield  {author} {\bibinfo {author} {\bibfnamefont {C.}~\bibnamefont {Cedzich}}\ and\ \bibinfo {author} {\bibfnamefont {A.~H.}\ \bibnamefont {Werner}},\ }\bibfield  {title} {\bibinfo {title} {{Anderson localization for electric quantum walks and skew-shift CMV matrices}},\ }\href {https://doi.org/10.1007/s00220-021-04204-w} {\bibfield  {journal} {\bibinfo  {journal} {Commun. Math. Phys.}\ }\textbf {\bibinfo {volume} {387}},\ \bibinfo {pages} {1257} (\bibinfo {year} {2021})},\ \Eprint {https://arxiv.org/abs/1906.11931} {arXiv:1906.11931} \BibitemShut {NoStop}%
\bibitem [{\citenamefont {Zhu}(2024)}]{Zhu:2021}%
  \BibitemOpen
  \bibfield  {author} {\bibinfo {author} {\bibfnamefont {X.}~\bibnamefont {Zhu}},\ }\bibfield  {title} {\bibinfo {title} {{Localization for random CMV matrices}},\ }\href {https://doi.org/10.1016/j.jat.2023.106008} {\bibfield  {journal} {\bibinfo  {journal} {J. Approx. Theory}\ }\textbf {\bibinfo {volume} {298}},\ \bibinfo {pages} {106008} (\bibinfo {year} {2024})},\ \Eprint {https://arxiv.org/abs/2110.11386} {arXiv:2110.11386} \BibitemShut {NoStop}%
\bibitem [{\citenamefont {Haake}\ \emph {et~al.}(1987)\citenamefont {Haake}, \citenamefont {Ku\'{s}},\ and\ \citenamefont {Scharf}}]{Haake:1987}%
  \BibitemOpen
  \bibfield  {author} {\bibinfo {author} {\bibfnamefont {F.}~\bibnamefont {Haake}}, \bibinfo {author} {\bibfnamefont {M.}~\bibnamefont {Ku\'{s}}},\ and\ \bibinfo {author} {\bibfnamefont {R.}~\bibnamefont {Scharf}},\ }\bibfield  {title} {\bibinfo {title} {{Classical and quantum chaos for a kicked top}},\ }\href {https://doi.org/10.1007/BF01303727} {\bibfield  {journal} {\bibinfo  {journal} {Z. Physik B}\ }\textbf {\bibinfo {volume} {65}},\ \bibinfo {pages} {381} (\bibinfo {year} {1987})}\BibitemShut {NoStop}%
\bibitem [{\citenamefont {Izrailev}(1990)}]{Izrailev:1990}%
  \BibitemOpen
  \bibfield  {author} {\bibinfo {author} {\bibfnamefont {F.~M.}\ \bibnamefont {Izrailev}},\ }\bibfield  {title} {\bibinfo {title} {{Simple models of quantum chaos: Spectrum and eigenfunctions}},\ }\href {https://doi.org/10.1016/0370-1573(90)90067-C} {\bibfield  {journal} {\bibinfo  {journal} {Phys. Rept.}\ }\textbf {\bibinfo {volume} {196}},\ \bibinfo {pages} {299} (\bibinfo {year} {1990})}\BibitemShut {NoStop}%
\bibitem [{\citenamefont {Rozenbaum}\ \emph {et~al.}(2017)\citenamefont {Rozenbaum}, \citenamefont {Ganeshan},\ and\ \citenamefont {Galitski}}]{Rozenbaum:2016}%
  \BibitemOpen
  \bibfield  {author} {\bibinfo {author} {\bibfnamefont {E.~B.}\ \bibnamefont {Rozenbaum}}, \bibinfo {author} {\bibfnamefont {S.}~\bibnamefont {Ganeshan}},\ and\ \bibinfo {author} {\bibfnamefont {V.}~\bibnamefont {Galitski}},\ }\bibfield  {title} {\bibinfo {title} {{Lyapunov exponent and out-of-time-ordered correlator\textquoteright{}s growth rate in a chaotic system}},\ }\href {https://doi.org/10.1103/PhysRevLett.118.086801} {\bibfield  {journal} {\bibinfo  {journal} {Phys. Rev. Lett.}\ }\textbf {\bibinfo {volume} {118}},\ \bibinfo {pages} {086801} (\bibinfo {year} {2017})},\ \Eprint {https://arxiv.org/abs/1609.01707} {arXiv:1609.01707} \BibitemShut {NoStop}%
\bibitem [{\citenamefont {Vallini}\ and\ \citenamefont {Pappalardi}()}]{Vallini:2024}%
  \BibitemOpen
  \bibfield  {author} {\bibinfo {author} {\bibfnamefont {E.}~\bibnamefont {Vallini}}\ and\ \bibinfo {author} {\bibfnamefont {S.}~\bibnamefont {Pappalardi}},\ }\bibfield  {title} {\bibinfo {title} {{Long-time freeness in the kicked top}},\ }\Eprint {https://arxiv.org/abs/2411.12050} {arXiv:2411.12050} \BibitemShut {NoStop}%
\bibitem [{\citenamefont {Chaudhury}\ \emph {et~al.}(2009)\citenamefont {Chaudhury}, \citenamefont {Smith}, \citenamefont {Anderson}, \citenamefont {Ghose},\ and\ \citenamefont {Jessen}}]{Chaudhury:2009}%
  \BibitemOpen
  \bibfield  {author} {\bibinfo {author} {\bibfnamefont {S.}~\bibnamefont {Chaudhury}}, \bibinfo {author} {\bibfnamefont {A.}~\bibnamefont {Smith}}, \bibinfo {author} {\bibfnamefont {B.~E.}\ \bibnamefont {Anderson}}, \bibinfo {author} {\bibfnamefont {S.}~\bibnamefont {Ghose}},\ and\ \bibinfo {author} {\bibfnamefont {P.~S.}\ \bibnamefont {Jessen}},\ }\bibfield  {title} {\bibinfo {title} {{Quantum signatures of chaos in a kicked top}},\ }\href {https://doi.org/10.1038/nature08396} {\bibfield  {journal} {\bibinfo  {journal} {Nature}\ }\textbf {\bibinfo {volume} {461}},\ \bibinfo {pages} {768} (\bibinfo {year} {2009})}\BibitemShut {NoStop}%
\bibitem [{\citenamefont {Neill}\ \emph {et~al.}(2016)\citenamefont {Neill} \emph {et~al.}}]{Neill:2016}%
  \BibitemOpen
  \bibfield  {author} {\bibinfo {author} {\bibfnamefont {C.}~\bibnamefont {Neill}} \emph {et~al.},\ }\bibfield  {title} {\bibinfo {title} {{Ergodic dynamics and thermalization in an isolated quantum system}},\ }\href {https://doi.org/10.1038/nphys3830} {\bibfield  {journal} {\bibinfo  {journal} {Nature Phys.}\ }\textbf {\bibinfo {volume} {12}},\ \bibinfo {pages} {1037} (\bibinfo {year} {2016})},\ \Eprint {https://arxiv.org/abs/1601.00600} {arXiv:1601.00600} \BibitemShut {NoStop}%
\bibitem [{Note8()}]{Note8}%
  \BibitemOpen
  \bibinfo {note} {This approach allows us to get cleaner results for small Hilbert space dimensions, but superfluous for $d > 10^3$.}\BibitemShut {Stop}%
\bibitem [{\citenamefont {Berry}(1977)}]{Berry:1977b}%
  \BibitemOpen
  \bibfield  {author} {\bibinfo {author} {\bibfnamefont {M.~V.}\ \bibnamefont {Berry}},\ }\bibfield  {title} {\bibinfo {title} {{Semi-classical mechanics in phase space: A study of Wigner’s function}},\ }\href {https://doi.org/10.1098/rsta.1977.0145} {\bibfield  {journal} {\bibinfo  {journal} {Philos. Trans. R. Soc. A}\ }\textbf {\bibinfo {volume} {287}},\ \bibinfo {pages} {237} (\bibinfo {year} {1977})}\BibitemShut {NoStop}%
\bibitem [{\citenamefont {Prosen}(2000)}]{Prozen:2000}%
  \BibitemOpen
  \bibfield  {author} {\bibinfo {author} {\bibfnamefont {T.}~\bibnamefont {Prosen}},\ }\bibfield  {title} {\bibinfo {title} {{Exact time-correlation functions of quantum Ising chain in a kicking transversal magnetic field: spectral analysis of the adjoint propagator in Heisenberg picture}},\ }\href {https://doi.org/10.1143/PTPS.139.191} {\bibfield  {journal} {\bibinfo  {journal} {Prog. Theor. Phys. Suppl.}\ }\textbf {\bibinfo {volume} {139}},\ \bibinfo {pages} {191} (\bibinfo {year} {2000})}\BibitemShut {NoStop}%
\bibitem [{\citenamefont {Prosen}(2002)}]{Prozen:2002}%
  \BibitemOpen
  \bibfield  {author} {\bibinfo {author} {\bibfnamefont {T.}~\bibnamefont {Prosen}},\ }\bibfield  {title} {\bibinfo {title} {{On general relation between quantum ergodicity and fidelity of quantum dynamics}},\ }\href {https://doi.org/10.1103/PhysRevE.65.036208} {\bibfield  {journal} {\bibinfo  {journal} {Phys. Rev. E}\ }\textbf {\bibinfo {volume} {65}},\ \bibinfo {pages} {036208} (\bibinfo {year} {2002})},\ \Eprint {https://arxiv.org/abs/quant-ph/0106149} {arXiv:quant-ph/0106149} \BibitemShut {NoStop}%
\bibitem [{\citenamefont {Pineda}\ and\ \citenamefont {Prosen}(2007)}]{Prozen:2007}%
  \BibitemOpen
  \bibfield  {author} {\bibinfo {author} {\bibfnamefont {C.}~\bibnamefont {Pineda}}\ and\ \bibinfo {author} {\bibfnamefont {T.}~\bibnamefont {Prosen}},\ }\bibfield  {title} {\bibinfo {title} {{Non-universal level statistics in a chaotic quantum spin chain}},\ }\href {https://doi.org/10.1103/PhysRevE.76.061127} {\bibfield  {journal} {\bibinfo  {journal} {Phys. Rev. E}\ }\textbf {\bibinfo {volume} {76}},\ \bibinfo {pages} {061127} (\bibinfo {year} {2007})},\ \Eprint {https://arxiv.org/abs/quant-ph/0702164} {arXiv:quant-ph/0702164} \BibitemShut {NoStop}%
\bibitem [{\citenamefont {Bertini}\ \emph {et~al.}(2018)\citenamefont {Bertini}, \citenamefont {Kos},\ and\ \citenamefont {Prosen}}]{Prozen:2018}%
  \BibitemOpen
  \bibfield  {author} {\bibinfo {author} {\bibfnamefont {B.}~\bibnamefont {Bertini}}, \bibinfo {author} {\bibfnamefont {P.}~\bibnamefont {Kos}},\ and\ \bibinfo {author} {\bibfnamefont {T.}~\bibnamefont {Prosen}},\ }\bibfield  {title} {\bibinfo {title} {{Exact spectral form factor in a minimal model of many-body quantum chaos}},\ }\href {https://doi.org/10.1103/PhysRevLett.121.264101} {\bibfield  {journal} {\bibinfo  {journal} {Phys. Rev. Lett.}\ }\textbf {\bibinfo {volume} {121}},\ \bibinfo {pages} {264101} (\bibinfo {year} {2018})},\ \Eprint {https://arxiv.org/abs/1805.00931} {arXiv:1805.00931} \BibitemShut {NoStop}%
\bibitem [{\citenamefont {Bertini}\ \emph {et~al.}(2019{\natexlab{a}})\citenamefont {Bertini}, \citenamefont {Kos},\ and\ \citenamefont {Prosen}}]{Prozen:2019}%
  \BibitemOpen
  \bibfield  {author} {\bibinfo {author} {\bibfnamefont {B.}~\bibnamefont {Bertini}}, \bibinfo {author} {\bibfnamefont {P.}~\bibnamefont {Kos}},\ and\ \bibinfo {author} {\bibfnamefont {T.}~\bibnamefont {Prosen}},\ }\bibfield  {title} {\bibinfo {title} {{Entanglement spreading in a minimal model of maximal many-body quantum chaos}},\ }\href {https://doi.org/10.1103/PhysRevX.9.021033} {\bibfield  {journal} {\bibinfo  {journal} {Phys. Rev. X}\ }\textbf {\bibinfo {volume} {9}},\ \bibinfo {pages} {021033} (\bibinfo {year} {2019}{\natexlab{a}})},\ \Eprint {https://arxiv.org/abs/1812.05090} {arXiv:1812.05090} \BibitemShut {NoStop}%
\bibitem [{\citenamefont {Akila}\ \emph {et~al.}(2016)\citenamefont {Akila}, \citenamefont {Waltner}, \citenamefont {Gutkin},\ and\ \citenamefont {Guhr}}]{Akila:2016}%
  \BibitemOpen
  \bibfield  {author} {\bibinfo {author} {\bibfnamefont {M.}~\bibnamefont {Akila}}, \bibinfo {author} {\bibfnamefont {D.}~\bibnamefont {Waltner}}, \bibinfo {author} {\bibfnamefont {B.}~\bibnamefont {Gutkin}},\ and\ \bibinfo {author} {\bibfnamefont {T.}~\bibnamefont {Guhr}},\ }\bibfield  {title} {\bibinfo {title} {{Particle-time duality in the kicked Ising spin chain}},\ }\href {https://doi.org/10.1088/1751-8113/49/37/375101} {\bibfield  {journal} {\bibinfo  {journal} {J. Phys. A: Math. Theor.}\ }\textbf {\bibinfo {volume} {49}},\ \bibinfo {pages} {375101} (\bibinfo {year} {2016})},\ \Eprint {https://arxiv.org/abs/1602.07130} {arXiv:1602.07130} \BibitemShut {NoStop}%
\bibitem [{Note9()}]{Note9}%
  \BibitemOpen
  \bibinfo {note} {The simplest example is quantized system with a mixed phase space, where regular and chaotic regions coexist.}\BibitemShut {Stop}%
\bibitem [{\citenamefont {Fisher}\ \emph {et~al.}(2023)\citenamefont {Fisher}, \citenamefont {Khemani}, \citenamefont {Nahum},\ and\ \citenamefont {Vijay}}]{Fisher:2022}%
  \BibitemOpen
  \bibfield  {author} {\bibinfo {author} {\bibfnamefont {M.~P.~A.}\ \bibnamefont {Fisher}}, \bibinfo {author} {\bibfnamefont {V.}~\bibnamefont {Khemani}}, \bibinfo {author} {\bibfnamefont {A.}~\bibnamefont {Nahum}},\ and\ \bibinfo {author} {\bibfnamefont {S.}~\bibnamefont {Vijay}},\ }\bibfield  {title} {\bibinfo {title} {{Random quantum circuits}},\ }\href {https://doi.org/10.1146/annurev-conmatphys-031720-030658} {\bibfield  {journal} {\bibinfo  {journal} {Ann. Rev. Condensed Matter Phys.}\ }\textbf {\bibinfo {volume} {14}},\ \bibinfo {pages} {335} (\bibinfo {year} {2023})},\ \Eprint {https://arxiv.org/abs/2207.14280} {arXiv:2207.14280} \BibitemShut {NoStop}%
\bibitem [{\citenamefont {Khemani}\ \emph {et~al.}(2018)\citenamefont {Khemani}, \citenamefont {Vishwanath},\ and\ \citenamefont {Huse}}]{Khemani:2017}%
  \BibitemOpen
  \bibfield  {author} {\bibinfo {author} {\bibfnamefont {V.}~\bibnamefont {Khemani}}, \bibinfo {author} {\bibfnamefont {A.}~\bibnamefont {Vishwanath}},\ and\ \bibinfo {author} {\bibfnamefont {D.~A.}\ \bibnamefont {Huse}},\ }\bibfield  {title} {\bibinfo {title} {{Operator spreading and the emergence of dissipation in unitary dynamics with conservation laws}},\ }\href {https://doi.org/10.1103/PhysRevX.8.031057} {\bibfield  {journal} {\bibinfo  {journal} {Phys. Rev. X}\ }\textbf {\bibinfo {volume} {8}},\ \bibinfo {pages} {031057} (\bibinfo {year} {2018})},\ \Eprint {https://arxiv.org/abs/1710.09835} {arXiv:1710.09835} \BibitemShut {NoStop}%
\bibitem [{\citenamefont {Arute}\ \emph {et~al.}(2019)\citenamefont {Arute} \emph {et~al.}}]{Arute:2019}%
  \BibitemOpen
  \bibfield  {author} {\bibinfo {author} {\bibfnamefont {F.}~\bibnamefont {Arute}} \emph {et~al.},\ }\bibfield  {title} {\bibinfo {title} {{Quantum supremacy using a programmable superconducting processor}},\ }\href {https://doi.org/10.1038/s41586-019-1666-5} {\bibfield  {journal} {\bibinfo  {journal} {Nature}\ }\textbf {\bibinfo {volume} {574}},\ \bibinfo {pages} {505} (\bibinfo {year} {2019})},\ \Eprint {https://arxiv.org/abs/1910.11333} {arXiv:1910.11333} \BibitemShut {NoStop}%
\bibitem [{\citenamefont {Bertini}\ \emph {et~al.}(2019{\natexlab{b}})\citenamefont {Bertini}, \citenamefont {Kos},\ and\ \citenamefont {Prosen}}]{Bertini:2019}%
  \BibitemOpen
  \bibfield  {author} {\bibinfo {author} {\bibfnamefont {B.}~\bibnamefont {Bertini}}, \bibinfo {author} {\bibfnamefont {P.}~\bibnamefont {Kos}},\ and\ \bibinfo {author} {\bibfnamefont {T.}~\bibnamefont {Prosen}},\ }\bibfield  {title} {\bibinfo {title} {{Exact correlation functions for dual-unitary lattice models in 1+1 dimensions}},\ }\href {https://doi.org/10.1103/PhysRevLett.123.210601} {\bibfield  {journal} {\bibinfo  {journal} {Phys. Rev. Lett.}\ }\textbf {\bibinfo {volume} {123}},\ \bibinfo {pages} {210601} (\bibinfo {year} {2019}{\natexlab{b}})},\ \Eprint {https://arxiv.org/abs/1904.02140} {arXiv:1904.02140} \BibitemShut {NoStop}%
\bibitem [{\citenamefont {Fritzsch}\ and\ \citenamefont {Prosen}(2021)}]{Fritzsch:2021}%
  \BibitemOpen
  \bibfield  {author} {\bibinfo {author} {\bibfnamefont {F.}~\bibnamefont {Fritzsch}}\ and\ \bibinfo {author} {\bibfnamefont {T.}~\bibnamefont {Prosen}},\ }\bibfield  {title} {\bibinfo {title} {{Eigenstate thermalization in dual-unitary quantum circuits: Asymptotics of spectral functions}},\ }\href {https://doi.org/10.1103/PhysRevE.103.062133} {\bibfield  {journal} {\bibinfo  {journal} {Phys. Rev. E}\ }\textbf {\bibinfo {volume} {103}},\ \bibinfo {pages} {062133} (\bibinfo {year} {2021})},\ \Eprint {https://arxiv.org/abs/2103.11694} {arXiv:2103.11694} \BibitemShut {NoStop}%
\bibitem [{\citenamefont {Gopalakrishnan}\ and\ \citenamefont {Lamacraft}(2019)}]{Gopalakrishnan:2019}%
  \BibitemOpen
  \bibfield  {author} {\bibinfo {author} {\bibfnamefont {S.}~\bibnamefont {Gopalakrishnan}}\ and\ \bibinfo {author} {\bibfnamefont {A.}~\bibnamefont {Lamacraft}},\ }\bibfield  {title} {\bibinfo {title} {{Unitary circuits of finite depth and infinite width from quantum channels}},\ }\href {https://doi.org/10.1103/PhysRevB.100.064309} {\bibfield  {journal} {\bibinfo  {journal} {Phys. Rev. B}\ }\textbf {\bibinfo {volume} {100}},\ \bibinfo {pages} {064309} (\bibinfo {year} {2019})},\ \Eprint {https://arxiv.org/abs/1903.11611} {arXiv:1903.11611} \BibitemShut {NoStop}%
\bibitem [{\citenamefont {Lerose}\ \emph {et~al.}(2021)\citenamefont {Lerose}, \citenamefont {Sonner},\ and\ \citenamefont {Abanin}}]{Lerose:2020}%
  \BibitemOpen
  \bibfield  {author} {\bibinfo {author} {\bibfnamefont {A.}~\bibnamefont {Lerose}}, \bibinfo {author} {\bibfnamefont {M.}~\bibnamefont {Sonner}},\ and\ \bibinfo {author} {\bibfnamefont {D.~A.}\ \bibnamefont {Abanin}},\ }\bibfield  {title} {\bibinfo {title} {{Influence matrix approach to many-body Floquet dynamics}},\ }\href {https://doi.org/10.1103/PhysRevX.11.021040} {\bibfield  {journal} {\bibinfo  {journal} {Phys. Rev. X}\ }\textbf {\bibinfo {volume} {11}},\ \bibinfo {pages} {021040} (\bibinfo {year} {2021})},\ \Eprint {https://arxiv.org/abs/2009.10105} {arXiv:2009.10105} \BibitemShut {NoStop}%
\bibitem [{\citenamefont {Po}\ \emph {et~al.}(2016)\citenamefont {Po}, \citenamefont {Fidkowski}, \citenamefont {Morimoto}, \citenamefont {Potter},\ and\ \citenamefont {Vishwanath}}]{Po:2016}%
  \BibitemOpen
  \bibfield  {author} {\bibinfo {author} {\bibfnamefont {H.~C.}\ \bibnamefont {Po}}, \bibinfo {author} {\bibfnamefont {L.}~\bibnamefont {Fidkowski}}, \bibinfo {author} {\bibfnamefont {T.}~\bibnamefont {Morimoto}}, \bibinfo {author} {\bibfnamefont {A.~C.}\ \bibnamefont {Potter}},\ and\ \bibinfo {author} {\bibfnamefont {A.}~\bibnamefont {Vishwanath}},\ }\bibfield  {title} {\bibinfo {title} {{Chiral Floquet phases of many-body localized bosons}},\ }\href {https://doi.org/10.1103/PhysRevX.6.041070} {\bibfield  {journal} {\bibinfo  {journal} {Phys. Rev. X}\ }\textbf {\bibinfo {volume} {6}},\ \bibinfo {pages} {041070} (\bibinfo {year} {2016})},\ \Eprint {https://arxiv.org/abs/1609.00006} {arXiv:1609.00006} \BibitemShut {NoStop}%
\bibitem [{\citenamefont {Nathan}\ \emph {et~al.}(2019)\citenamefont {Nathan}, \citenamefont {Abanin}, \citenamefont {Berg}, \citenamefont {Lindner},\ and\ \citenamefont {Rudner}}]{Nathan:2019}%
  \BibitemOpen
  \bibfield  {author} {\bibinfo {author} {\bibfnamefont {F.}~\bibnamefont {Nathan}}, \bibinfo {author} {\bibfnamefont {D.}~\bibnamefont {Abanin}}, \bibinfo {author} {\bibfnamefont {E.}~\bibnamefont {Berg}}, \bibinfo {author} {\bibfnamefont {N.~H.}\ \bibnamefont {Lindner}},\ and\ \bibinfo {author} {\bibfnamefont {M.~S.}\ \bibnamefont {Rudner}},\ }\bibfield  {title} {\bibinfo {title} {{Anomalous Floquet insulators}},\ }\href {https://doi.org/10.1103/PhysRevB.99.195133} {\bibfield  {journal} {\bibinfo  {journal} {Phys. Rev. B}\ }\textbf {\bibinfo {volume} {99}},\ \bibinfo {pages} {195133} (\bibinfo {year} {2019})},\ \Eprint {https://arxiv.org/abs/1712.02789} {arXiv:1712.02789} \BibitemShut {NoStop}%
\bibitem [{\citenamefont {Wilczek}(2012)}]{Wilczek:2012}%
  \BibitemOpen
  \bibfield  {author} {\bibinfo {author} {\bibfnamefont {F.}~\bibnamefont {Wilczek}},\ }\bibfield  {title} {\bibinfo {title} {{Quantum time crystals}},\ }\href {https://doi.org/10.1103/PhysRevLett.109.160401} {\bibfield  {journal} {\bibinfo  {journal} {Phys. Rev. Lett.}\ }\textbf {\bibinfo {volume} {109}},\ \bibinfo {pages} {160401} (\bibinfo {year} {2012})},\ \Eprint {https://arxiv.org/abs/1202.2539} {arXiv:1202.2539} \BibitemShut {NoStop}%
\bibitem [{\citenamefont {Khemani}\ \emph {et~al.}(2016)\citenamefont {Khemani}, \citenamefont {Lazarides}, \citenamefont {Moessner},\ and\ \citenamefont {Sondhi}}]{Khemani:2016}%
  \BibitemOpen
  \bibfield  {author} {\bibinfo {author} {\bibfnamefont {V.}~\bibnamefont {Khemani}}, \bibinfo {author} {\bibfnamefont {A.}~\bibnamefont {Lazarides}}, \bibinfo {author} {\bibfnamefont {R.}~\bibnamefont {Moessner}},\ and\ \bibinfo {author} {\bibfnamefont {S.~L.}\ \bibnamefont {Sondhi}},\ }\bibfield  {title} {\bibinfo {title} {{Phase structure of driven quantum systems}},\ }\href {https://doi.org/10.1103/PhysRevLett.116.250401} {\bibfield  {journal} {\bibinfo  {journal} {Phys. Rev. Lett.}\ }\textbf {\bibinfo {volume} {116}},\ \bibinfo {pages} {250401} (\bibinfo {year} {2016})},\ \Eprint {https://arxiv.org/abs/1508.03344} {arXiv:1508.03344} \BibitemShut {NoStop}%
\bibitem [{\citenamefont {Else}\ \emph {et~al.}(2016)\citenamefont {Else}, \citenamefont {Bauer},\ and\ \citenamefont {Nayak}}]{Else:2016}%
  \BibitemOpen
  \bibfield  {author} {\bibinfo {author} {\bibfnamefont {D.~V.}\ \bibnamefont {Else}}, \bibinfo {author} {\bibfnamefont {B.}~\bibnamefont {Bauer}},\ and\ \bibinfo {author} {\bibfnamefont {C.}~\bibnamefont {Nayak}},\ }\bibfield  {title} {\bibinfo {title} {{Floquet time crystals}},\ }\href {https://doi.org/10.1103/PhysRevLett.117.090402} {\bibfield  {journal} {\bibinfo  {journal} {Phys. Rev. Lett.}\ }\textbf {\bibinfo {volume} {117}},\ \bibinfo {pages} {090402} (\bibinfo {year} {2016})},\ \Eprint {https://arxiv.org/abs/1603.08001} {arXiv:1603.08001} \BibitemShut {NoStop}%
\bibitem [{\citenamefont {Choi}\ \emph {et~al.}(2017)\citenamefont {Choi} \emph {et~al.}}]{Choi:2017}%
  \BibitemOpen
  \bibfield  {author} {\bibinfo {author} {\bibfnamefont {S.}~\bibnamefont {Choi}} \emph {et~al.},\ }\bibfield  {title} {\bibinfo {title} {{Observation of discrete time-crystalline order in a disordered dipolar many-body system}},\ }\href {https://doi.org/10.1038/nature21426} {\bibfield  {journal} {\bibinfo  {journal} {Nature}\ }\textbf {\bibinfo {volume} {543}},\ \bibinfo {pages} {221} (\bibinfo {year} {2017})},\ \Eprint {https://arxiv.org/abs/1610.08057} {arXiv:1610.08057} \BibitemShut {NoStop}%
\bibitem [{\citenamefont {Zhang}\ \emph {et~al.}(2017)\citenamefont {Zhang} \emph {et~al.}}]{Zhang:2017}%
  \BibitemOpen
  \bibfield  {author} {\bibinfo {author} {\bibfnamefont {J.}~\bibnamefont {Zhang}} \emph {et~al.},\ }\bibfield  {title} {\bibinfo {title} {{Observation of a discrete time crystal}},\ }\href {https://doi.org/10.1038/nature21413} {\bibfield  {journal} {\bibinfo  {journal} {Nature}\ }\textbf {\bibinfo {volume} {543}},\ \bibinfo {pages} {217} (\bibinfo {year} {2017})},\ \Eprint {https://arxiv.org/abs/1609.08684} {arXiv:1609.08684} \BibitemShut {NoStop}%
\bibitem [{\citenamefont {Derevyagin}\ \emph {et~al.}(2012)\citenamefont {Derevyagin}, \citenamefont {Vinet},\ and\ \citenamefont {Zhedanov}}]{DVG}%
  \BibitemOpen
  \bibfield  {author} {\bibinfo {author} {\bibfnamefont {M.}~\bibnamefont {Derevyagin}}, \bibinfo {author} {\bibfnamefont {L.}~\bibnamefont {Vinet}},\ and\ \bibinfo {author} {\bibfnamefont {A.}~\bibnamefont {Zhedanov}},\ }\bibfield  {title} {\bibinfo {title} {{CMV matrices and little and big $-1$ Jacobi polynomials}},\ }\href {https://doi.org/10.1007/s00365-012-9164-0} {\bibfield  {journal} {\bibinfo  {journal} {Constr. Approx.}\ }\textbf {\bibinfo {volume} {36}},\ \bibinfo {pages} {513} (\bibinfo {year} {2012})},\ \Eprint {https://arxiv.org/abs/1108.3535} {arXiv:1108.3535} \BibitemShut {NoStop}%
\bibitem [{\citenamefont {Cantero}\ \emph {et~al.}(2021)\citenamefont {Cantero}, \citenamefont {Marcell{\'a}n}, \citenamefont {Moral},\ and\ \citenamefont {Velazquez}}]{CMV:2020}%
  \BibitemOpen
  \bibfield  {author} {\bibinfo {author} {\bibfnamefont {M.~J.}\ \bibnamefont {Cantero}}, \bibinfo {author} {\bibfnamefont {F.}~\bibnamefont {Marcell{\'a}n}}, \bibinfo {author} {\bibfnamefont {L.}~\bibnamefont {Moral}},\ and\ \bibinfo {author} {\bibfnamefont {L.}~\bibnamefont {Velazquez}},\ }\bibfield  {title} {\bibinfo {title} {{A CMV connection between orthogonal polynomials on the unit circle and the real line}},\ }\href {https://doi.org/10.1016/j.jat.2021.105579} {\bibfield  {journal} {\bibinfo  {journal} {J. Approx. Theory}\ }\textbf {\bibinfo {volume} {266}},\ \bibinfo {pages} {105579} (\bibinfo {year} {2021})},\ \Eprint {https://arxiv.org/abs/2005.09772} {arXiv:2005.09772} \BibitemShut {NoStop}%
\end{thebibliography}%

\end{document}